\documentclass[useAMS,usenatbib]{mn2e}

\usepackage{latexsym}
\usepackage{amssymb}
\usepackage{graphicx}

\usepackage{subfigure}

\newcommand{\detg}{{\sqrt{-g}}}
\newcommand{\del}{{\partial}}

\newcommand{\pc}{{\rm pc}}

\newcommand{\msun}{{\rm M_{\odot}}}

\newcommand{\dF}{{^{^*}\!\!F}}
\newcommand{\alf}{Alfv\'en}

\newcommand{\bB}{{\bf B}}

\newcommand{\bva}{{\bf v}_{\rm A}}
\newcommand{\va}{{\bf v}_{\rm A}}
\newcommand{\kdv}{({\bf{k}} \cdot {\bf v}_{\rm A})}

\newcommand{\km}{{\rm\,km}}
\newcommand{\yr}{{\rm\,yr}}

\newcommand{\cm}{{\rm\,cm}}

\newcommand{\ergps}{{\rm\,erg~s^{-1}}}

\newcommand{\sE}{{\mathcal{E}}}
\newcommand{\IEDEN}{{u_g}}

\def\fps@figure{bp}%
\def\fps@table{bp}%
\def\fps@plate{bp}%
\def\eps@scaling{1.0}%
\newcommand\epsscale[1]{\gdef\eps@scaling{#1}}%
\newcommand\plotone[1]{%
 \centering
 \leavevmode
 \includegraphics[width={\eps@scaling\columnwidth}]{#1}%
}%
\newcommand\plottwo[2]{%
 \centering
 \leavevmode
 \columnwidth=.45\columnwidth
 \includegraphics[width={\eps@scaling\columnwidth}]{#1}%
 \hfil
 \includegraphics[width={\eps@scaling\columnwidth}]{#2}%
}%
\newcommand\plotfiddle[7]{%
 \centering
 \leavevmode
 \vbox\@to#2{\rule{\z@}{#2}}%
 \includegraphics[%
  scale=#4,
  angle=#3,
  origin=c
 ]{#1}%
}%

\newcommand\araa{\rmfamily{ARA\&A}}%
\newcommand\apj{\rmfamily{ApJ}}%
\newcommand\apjl{\rmfamily{ApJ}}%
\newcommand\apjs{\rmfamily{ApJS}}%
\newcommand\apss{\rmfamily{Ap\&SS}}%
\newcommand\aap{\rmfamily{A\&A}}%
\newcommand\mnras{\rmfamily{MNRAS}}%
\newcommand\prd{\rmfamily{Phys.~Rev.~D}}%
\newcommand\pasp{\rmfamily{PASP}}%
\newcommand\pasj{\rmfamily{PASJ}}%
\newcommand\nat{\rmfamily{Nature}}%
\newcommand\jgr{\rmfamily{J.~Geophys.~Res.}}%
\title[GRMHD Simulations of Poynting-dominated Jets]{General Relativistic Magnetohydrodynamic Simulations of Jet Formation and Large-Scale Propagation from Black Hole Accretion Systems}

\author[Jonathan C. McKinney]{Jonathan
  C. McKinney\thanks{E-mail:jmckinney@cfa.harvard.edu High Res. Figures:  {\rm http://rainman.astro.uiuc.edu/\textasciitilde jon/jetlarge.pdf}}\\ Institute for
  Theory and Computation, Harvard-Smithsonian Center for Astrophysics,
  60 Garden Street, MS 51, Cambridge, MA 02138, USA}

\begin{document}
\date{Accepted 2006 March 01.  Received 2006 February 25; in original form 2006 January 06}
\pagerange{\pageref{firstpage}--\pageref{lastpage}} \pubyear{2006}
\maketitle \label{firstpage}

\begin{abstract}

The formation and large-scale propagation of Poynting-dominated jets
produced by accreting, rapidly rotating black hole systems are
studied by numerically integrating the general relativistic
magnetohydrodynamic equations of motion to follow the
self-consistent interaction between accretion disks and black holes.
This study extends previous similar work by studying jets till
$t\approx 10^4GM/c^3$ out to $r\approx 10^4GM/c^2$, by which the jet
is super- fast magnetosonic and moves at a lab-frame bulk Lorentz
factor of $\Gamma\sim 10$ with a maximum terminal Lorentz factor of
$\Gamma_\infty\lesssim 10^3$. The radial structure of the
Poynting-dominated jet is piece-wise self-similar, and fits to flow
quantities along the field line are provided.  Beyond the
\alf~surface at $r\sim 10$--$100GM/c^2$, the jet becomes marginally
unstable to (at least) current-driven instabilities.  Such
instabilities drive shocks in the jet that limit the efficiency of
magnetic acceleration and collimation. These instabilities also
induce jet substructure with $3\lesssim\Gamma\lesssim 15$.   The jet
is shown to only marginally satisfy the necessary and sufficient
conditions for kink instability, so this may explain how
astrophysical jets can extend to large distances without completely
disrupting. At large distance, the jet angular structure is
Gaussian-like (or uniform within the core with sharp exponential
wings) with a half-opening angle of $\approx 5^\circ$ and there is
an extended component out to $\approx 27^\circ$. Unlike in some
hydrodynamic simulations, the environment is found to play a
negligible role in jet structure, acceleration, and collimation as
long as the ambient pressure of the surrounding medium is small
compared to the magnetic pressure in the jet.

\end{abstract}




\section{Introduction}\label{introduction}

Gamma-ray bursts (GRBs), active galactic nuclei (AGN), and black hole
x-ray binary systems exhibit relativistic jets that are believed to be
driven by accreting black holes (see, e.g.,
\citealt{rees82,begelman84,w93}).  Accreting black holes are the
highly efficient engines necessary to account for the observed
kinematics and energetics of such relativistic jets.  Understanding
systems powered by black hole accretion requires a general
relativistic magnetohydrodynamics (GRMHD) model of the bulk flow
dynamics near the black hole where the relativistic jet is launched.
The ideal MHD approximation has been shown to be a reasonably valid
model to describe the nonradiative dynamically important black hole
accretion physics associated with GRBs, AGN, and black hole x-ray
binary systems (see, e.g., \citealt{phi83,mckinney2004}).  This
approximation is the foundation of most studies of jets and winds.

We are concerned with explaining how relativistic jets are formed,
accelerate, collimate, and what is the resulting structure at large
radii.  The primarily difficulty has been to explain how observed jets
come to be so well-collimated with opening angles of a few degrees,
fast with bulk Lorentz factors of up to $\Gamma\sim 10^3$, and why
magnetized jets would not be disrupted by instabilities, such as the
kink instability.  It has not even been known whether the MHD
approximation was sufficient to understand the self-consistent
production of jets or their observed fast speeds and small opening
angles.

GRMHD numerical models of black hole accretion systems have
significantly progressed our understanding of relativistic jets.  Such
simulations have displayed primarily two types of ``jets,'' where one
is associated with the disk that is mass-loaded by disk material and
the other is associated directly with the black hole
\citep{dhk03,mg04,dv05a}.

GRMHD numerical models of winds from the inner part of the accretion
disk have found that the Lorentz factor only reaches $\Gamma<3$
\citep{dhk03,mg04,dv05a}. This includes GRMHD numerical models of GRBs
that showed a disk wind with a velocity of only $v\sim 0.3c$
\citep{mizuno04b}. Thus, disk winds may not explain jets with high
Lorentz factors unless additional physics, such as possibly efficient
neutrino-annihilation for GRBs or shock acceleration of particles in
general, is included.

The most astrophysically plausible mechanism to generate highly
relativistic jets is via the electromagnetic extraction of black hole
spin energy by the Blandford-Znajek (BZ) mechanism \citep{bz77}.  They
modelled the disk as a current sheet that sources the magnetic field
that couples to a rotating black hole, which spins down due to a
transfer of rotational energy of the space-time to the magnetic field.
However, their force-free model does not include matter, and so cannot
directly explain the Lorentz factor of observed jets. Also, the BZ
model does not self-consistently determine the current in the disk or
the field geometry that threads the black hole.  All these features
require the inclusion of matter, or the inertial effects of matter,
that would be present in an accretion disk.

GRMHD numerical models of black hole accretion have shown polar
regions with Poynting-dominated jets that are consistent with being
powered by the BZ effect \citep{mg04,kom05,mckinney2005a}.  They have
also shown that the magnetic field corresponding to the
Blandford-Znajek effect dominates all other magnetic field geometries
\citep{hirose04,mckinney2005a}.  In contrast to the disk wind that is
self-consistently mass-loaded by the disk itself, the
Poynting-dominated jet that is launched by the BZ effect has an
arbitrary Lorentz factor that is determined by the detailed
high-energy physics of the mass-loading of the jet, so low
mass-loading can explain highly relativistic jets (see, e.g.
\citealt{phi83,beskin97,punsly2001,lev05,mckinney2005a,mckinney2005b}).

We study a generic black hole accretion system in the GRMHD
approximation and consider its application to all black hole accretion
systems.  This unification scheme works because the GRMHD equations of
motion scale arbitrarily with the mass of the black hole and the mass
accretion rate.  Hence, a GRMHD model can be applied to any black hole
accretion system as a unified model, apart from important high-energy
effects that break this scale invariance.

The primary modifications to the GRMHD model come from radiative and
other high-energy effects.  As in any other study, we assume that the
results of a GRMHD model are interesting as long as such physics plays
a simple role in the bulk jet dynamics.  This role includes setting
the mass-loading of the Poynting jet, where in our GRMHD model this is
set to some fixed value.  Also, the jet speed may be slowly modified,
such as by Compton scattering or by synchrotron emission of the
internal energy.  In this case the radiative effects can be treated as
small corrections to the background state set by the GRMHD model.

This paper extends the work of \citet{mg04} by studying the
self-consistently generated Poynting-dominated jet as it propagates to
large distances.  A new version of the GRMHD code called HARM
\citep{gmt03} is used to integrate the equations of motion.  Unlike
prior work, the jet is shown to reach bulk Lorentz factors of
$\Gamma\sim 10$ and maximum terminal Lorentz factors with
$\Gamma_\infty\lesssim 10^3$ and collimate to narrow half-opening
angles of $\theta_j\sim 5^\circ$.  The simulated jet is sufficiently
fast and collimated to account for jet observations associated with
GRBs, AGN, or x-ray binary systems.

\subsection*{Outline of Paper}

In Section~\ref{mot}, we discuss the astrophysical systems to which
this study can be applied.  We discuss the direct or indirect
observations of jet speed, collimation, and structure, which are the
principle issues this paper attempts to address.

In Section~\ref{GRMHD}, we summarize the equations solved and the
units used in the paper.  Examples are given for how to apply our
results to any astrophysical system.

In Section~\ref{setup}, we present the setup of the numerical model,
such as the initial and boundary conditions.  The astrophysical
applicability of the model setup to GRBs, AGN, and x-ray binaries is
considered.

In Section~\ref{results}, we present the results of a fiducial
numerical model and describe the jet formation, structure, and
propagation to large distances.  Both the radial and angular
structures of the jet are studied and data fits to flow quantities are
given.  The jet characteristic structure, such as the formation of a
double transonic (trans-fast) flow, is discussed.

In Section~\ref{discussion}, we discuss the results and their
implications.  The results are compared to similar investigations, and
the limitations of the models presented are discussed.

In Section~\ref{conclusions}, we provide a short summary of the key
points.


\section{Astrophysical Motivation}\label{mot}

Observations associated with GRBs, AGN, and black hole x-ray binary
systems all show direct evidence of relativistic outflows. While AGN
and x-ray binaries show direct evidence of jet collimation, outflows
associated with GRBs are believed to be collimated based on achromatic
breaks in afterglow light curves.  However, the formation, structure,
acceleration, and collimation of relativistic jets are not yet
understood.  The following discussion outlines the motivation for
using a GRMHD model to study jets from GRBs, AGN, and x-ray binary
systems.

\subsection{GRBs}

Neutron stars and black holes are associated with the most violent of
post-Big Bang events: supernovae and some GRBs and maybe some x-ray
flashes (XRFs) (for a general review see \citealt{w93,wyhw00}).
Observations of a supernova light curve (SN2003dh) in the afterglow of
GRB 030329 suggest that at least some long-duration GRBs are
consistent with core-collapse events
\citep{stanek2003,kawabata2003,uemura2003, hjorth2003}.

Neutrino processes and magnetic fields are both important to
understand core-collapse.  In unraveling the mechanism by which
core-collapse supernovae explode, the details of the neutrino
transport model was at some point realized to be critical to whether a
supernova is produced in simulations \citep{Trans}. This has thus far
been interpreted to imply that highly accurate neutrino transport
physics is required, but this could also mean additional physics, such
as a magnetic field, could play a significant role. Indeed, all
core-collapse events may be powered by MHD processes rather than
neutrino processes \citep{lw70,sym84,ww86,dt92,khok99,awml03}.
Core-collapse involves shearing subject to the Balbus-Hawley
instability as in accretion disks \citep{awml03}, and this process
creates a dynamically important magnetic field during collapse.  All
core-collapse explosions are significantly polarised, asymmetric, and
often bi-polar indicating a strong role of rotation and a magnetic
field (see, e.g., \citealt{ww96,wyhw00,wang01,wang02,wang03}, and
references therein). Possible evidence for a magnetic dominated
outflow has been found in GRB 021206 \citep{cb03}, marginally
consistent with a magnetic outflow directly from the inner engine
\citep{lpb03}, although these observations remain
controversial. However, most core-collapse supernovae may still not
require magnetic fields \citep{woosley03}.

The key energy source for many GRB models is a black hole accretion
system since the efficiency is high for converting gravitational
binding energy to some type of emission.  Collapsar type models are
based upon the formation of a black hole during the core-collapse of
rapidly rotating massive stars.  The typical radius of the accretion
disk may determine the duration of long-duration GRBs
\citep{w93,pac98,mw99}.  An accretion disk is also formed as a result
of a neutron star or black hole collisions with another stellar object
\citep{narayan1992,narayan2001}.

GRBs are believed to be the result of an ultrarelativistic
jet. Indirect observational evidence of relativistic motion is
suggested by afterglow achromatic light breaks and the ``compactness
problem'' suggests GRB material must be ultrarelativistic with Lorentz
factor $\Gamma\gtrsim 100$ to emit the observed nonthermal
$\gamma$-rays (see, e.g., \citealt{piran2005}).  Direct observational
evidence for relativistic motion comes from radio scintillation of the
ISM \citep{goodman1997} and measurements of the afterglow emitting
region from GRB030329 \citep{taylor2004a,taylor2004b}.

Typical GRB jet models invoke either a hot neutrino-driven jet or a
cold Poynting flux-dominated jet.  While both mechanisms extract
comparable amounts of the accretion energy to power the jet
\citep{pwf99}, a neutrino-driven jet derives its energy from the
annihilation of neutrinos produced by the release of gravitational
binding energy, and consequently the jet is thermally accelerated.
However, strong outflows can be magnetically driven
\citep{br76,lovelace76,blandford76,bz77}.

\subsection{AGN}

AGN have long been believed to be powered by accretion onto
supermassive black holes \citep{z64,salpeter64}.  Observations of
MCG 6-30-15 show an iron line feature consistent with emission from
a relativistic disk with $v/c\sim 0.2$ \citep{tanaka1995,
fabianvaughan2002}, although the lack of a temporal correlation
between the continuum emission and iron-line emission may suggest
the basic fluorescence model is incorrect.  Alternatively, the
iron-line may be a jet-related feature \citep{elvis2000}.

AGN are observed to have jets with $\Gamma\sim 10$
\citep{up95,biretta99}, maybe even $\Gamma\sim 30$
\citep{brs94,gc01,jorstad01}, while some observations imply
$\Gamma\lesssim 200$ \citep{ghis93,kraw02,kono03}.  Some radio-quiet
AGN show evidence of weak jets \citep{ghis04}, which could be
explained as a disk wind and not require a rapidly rotating black hole
\citep{mg04}.  Observations imply the existence of a two-component jet
structure with a Poynting jet core and a dissipative surrounding
component \citep{ghis96,ghis05}.  The energy structure of the jet and
wind are important in understanding the feedback effect that controls
size of the black hole and may determine the $M-\sigma$ relation
\citep{springel2004,dimat05}.

\subsection{X-ray Binaries}

Long after their formation, neutron stars and black holes often
continue to produce outflows and jets \citep{mr99}.  These include
x-ray binaries (for a review see \citealt{lewin1995,mcclintock2003}),
neutron star as pulsars (for a review see \citealt{lorimer2001} on ms
pulsars and \citealt{tc99} on radio pulsars) and soft-gamma ray
repeaters (SGRs) \citep{td95,td96,kouv1999}.  In the case of x-ray
binaries, the companion star's stellar wind or Roche-lobe forms an
accretion disk.  Many x-ray binaries in their hard/low state (and
radio-loud AGN) show a correlation between the x-ray luminosity and
radio luminosity \citep{merloni2003}, which is consistent with radio
synchrotron emission from a jet and x-ray emission from a
geometrically thick, optically thin, Comptonizing disk.

Some black hole x-ray binaries have jets
\citep{mirabel1992,fender2003a}, such as GRS 1915+105 with apparently
superluminal motion \citep{mr94,mr99,fb04,kaiser04}.  However, it is
difficult to constrain the Lorentz factor, which in some ejection
events may be as large for x-ray binaries as for AGN with a Lorentz
factor of $\Gamma\lesssim 100$ \citep{fender2003b,mj06}.  The observed
broad, shifted, and asymmetric iron line from GRS 1915+105 is possible
evidence for a relativistic accretion disk \citep{martocchia02},
although this feature could be produced by a jet component as
suggested for AGN.

\citet{gd04} suggest that at least some systems, such as GRS 1915+105,
have slowly rotating black holes (but see \citealt{cui98}).  If this
is correct, then the BZ mechanism may not be responsible for these
jets.  A baryon-loaded disk wind with $\Gamma\lesssim 3$ can be
produced from a black hole accretion disk and not require a rapidly
rotating black hole \citep{mg04}.  Nonrelativistic outflows were found
even in viscous hydrodynamic simulations \citep{spb99,ia99,ia00,mg02}.

\section{Evolution Equations and Units}\label{GRMHD}

The model is that of a rotating black hole, described by the Kerr
metric, surrounded by an accretion disk.  The Kerr metric is written
in Kerr-Schild coordinates, such that the inner-radial computational
boundary can be placed inside the horizon and so out of causal contact
with the flow.  The Kerr metric in Kerr-Schild coordinates and the
Jacobian transformation to Boyer-Lindquist coordinates are given in
\citet{mg04}.

Boyer-Lindquist coordinates are not chosen because it is difficult to
avoid interactions between the inner-radial computational boundary and
the jet. The coordinate singularity at the event horizon in
Boyer-Lindquist can be avoided by placing the inner-radial
computational boundary outside the horizon.  However,
Poynting-dominated flows have waves that propagate outward even
arbitrarily close to the event horizon.  Using Boyer-Lindquist
coordinates can lead to excessive variability in the jet since the
ingoing superfast transition is not on the computational grid, and
then the details of the boundary condition can significantly influence
the jet.  Numerical models of viscous flows have historically had
related issues (see discussion in, e.g., \citealt{mg02}).

\subsection{GRMHD Equations of Motion}

The GRMHD notation follows \citet{mtw73}, hereafter MTW. A
single-component MHD approximation is assumed such that particle
number is conserved,
\begin{equation}
(\rho_0 u^\mu)_{;\mu} = 0 ,
\end{equation}
where $\rho_0$ is the rest-mass density and $u^\mu$ is the 4-velocity.
A 4-velocity with a spatial drift is introduced that is unique by
always being related to a physical observer for any space-time and has
well-behaved spatially interpolated values, which is useful for
numerical schemes.  This 4-velocity is
\begin{equation}
\tilde{u}^i \equiv u^i -\gamma \eta^i ,
\end{equation}
where $\gamma=-u^\alpha \eta_\alpha$.  This additional term represents
the spatial drift of the zero angular momentum (ZAMO) frame defined to
have a 4-velocity of $\eta_\mu = \{-\alpha,0,0,0\}$, where
$\alpha\equiv 1/\sqrt{-g^{tt}}$ and so $u^t = \gamma/\alpha$. One can
show that $\gamma=(1+q^2)^{1/2}$ with $q^2\equiv g_{ij}
\tilde{u}^i\tilde{u}^j$.

For a magnetized plasma, the conservation of energy-momentum
equations are
\begin{equation}\label{EOM}
{T^{\mu\nu}}_{;\nu} = \left(T^{\mu\nu}_{\rm MA} + T^{\mu\nu}_{\rm
EM}\right)_{;\nu} = 0.
\end{equation}
where $T^{\mu\nu}$ is the stress-energy tensor, which can be split
into a matter (MA) and electromagnetic (EM) part.  In the fluid
approximation
\begin{equation}
T^{\mu\nu}_{\rm MA} = (\rho_0 + u_g) u^\mu u^\nu + p_g P^{\mu\nu},
\end{equation}
with a relativistic ideal gas pressure $p_g=(\gamma-1) u_g$, where
$u_g$ is the internal energy density and the projection tensor is
$P^{\mu\nu} = g^{\mu\nu} + u^\mu u^\nu$, which projects any 4-vector
into the comoving frame (i.e. $P^{\nu\mu} u_\mu = 0$).

In terms of the Faraday (or electromagnetic field) tensor
($F^{\mu\nu}$),
\begin{equation}\label{tmunuem}
T^{\mu\nu}_{\rm EM} =
F^{\mu\gamma}{F^{\nu}}_{\gamma} -\frac{1}{4}g^{\mu\nu}
F^{\alpha\beta}F_{\alpha\beta},
\end{equation}
which is written in Heaviside-Lorentz units such that a factor of
$4\pi$ is absorbed into the definition of $F^{\mu\nu}$, where the
Gaussian unit value of the magnetic field is obtained by multiplying
the Heaviside-Lorentz value by $\sqrt{4\pi}$.  The induction equation
is given by the space components of ${\dF^{\mu\nu}}_{;\nu} = 0$, where
$\dF^{\mu\nu} = {1\over{2}} \epsilon^{\mu\nu\kappa\lambda}
F_{\kappa\lambda}$ is the dual of the Faraday tensor (Maxwell tensor),
and the time component gives the no-monopoles constraint.  Here
$\mathbf{\epsilon}$ is the Levi-Civita tensor, where
$\epsilon^{\mu\nu\lambda\delta} = -{1\over{\detg}}
[\mu\nu\lambda\delta]$ and $[\mu\nu\lambda\delta] $ is the completely
antisymmetric symbol.  The comoving electric field is defined as
\begin{equation}
e^\mu \equiv u_\nu F^{\mu\nu} = {1\over{2}}\epsilon^{\mu\nu k\lambda} u_\nu \dF_{\lambda k} = \eta j^\nu ,
\end{equation}
where $\eta$ corresponds to a scalar resistivity for a comoving
current density $j^\mu = J_\nu P^{\nu\mu}$.  The comoving
magnetic field is defined as
\begin{equation}\label{bcon}
b^\nu \equiv u_\mu \dF^{\mu\nu} = {1\over{2}}\epsilon^{\mu\nu
  k\lambda} u_\nu \dF_{\lambda k} .
\end{equation}
The ideal MHD approximation, $\eta = e^\mu=0$, is assumed, and so the
invariant $e^\mu b_\mu = 0$.  Since the Lorentz acceleration on a
particle is $f_l^\mu=q e^\mu$, then this implies that the Lorentz
force vanishes on a {\it particle} in the ideal MHD approximation.
Since $e^\nu u_\nu = b^\nu u_\nu = 0$, they each have only 3
independent components.  One can show that
\begin{equation}
\dF^{\mu\nu} = b^\mu u^\nu - b^\nu u^\mu ,
\end{equation}
and
\begin{equation}
F^{\mu\nu} = \epsilon^{\mu\nu\sigma\epsilon} u_\sigma b_\epsilon ,
\end{equation}
so that the electromagnetic part of the stress-energy tensor can be
written as
\begin{equation}
T^{\mu\nu}_{\rm EM} = {b^2\over{2}}(u^\mu u^\nu + P^{\mu\nu}) - b^\mu b^\nu .
\end{equation}
The other Maxwell equations, $J^\mu = {F^{\mu\nu}}_{;\nu}$, define the
current density, $J^\mu$, but are not needed in the ideal MHD
approximation for the evolution of the matter or the magnetic field.

For numerical simplicity, another set of field vectors are introduced,
such that $B^i \equiv \dF^{it}$ and $E_i \equiv F_{it}/\detg$.  The
two 4-vectors $e^\mu$ and $b^\mu$ and the 3-vectors $B^i$ and $E_i$
are just different ways of writing the independent components of the
Faraday or Maxwell tensors.  Equation (\ref{bcon}) implies $b^t = B^i
u_i$ and $b^i = (B^i + u^i b^t)/u^t$. Then the no-monopoles constraint
becomes
\begin{equation}
(\detg B^i)_{,i} = 0 ,
\end{equation}
and the magnetic induction equation becomes
\begin{eqnarray}\label{induct}
(\detg B^i)_{,t} & = & -(\detg(b^i u^j - b^j u^i ))_{,j} \nonumber\\
& = & -(\detg(B^i v^j - B^j v^i)_{,j} \nonumber\\
& = & -(\detg(\epsilon^{ijk}\varepsilon_k))_{,j} ,
\end{eqnarray}
where $v^i=u^i/u^t$, $E_i=\varepsilon_i=-\epsilon_{ijk} v^j
B^k=-\bf{v}\times\bf{B}$ is the electromotive force (EMF), and
$\epsilon^{ijk}$ is the spatial permutation tensor.  The above set of
equations are those that are solved.  A more complete discussion of
the relativistic MHD equations can be found in \cite{anile}.

\subsection{Floor Model}\label{floor}

In GRMHD numerical models of Poynting jets, the black hole's polar
region is always completely evacuated
\citep{mg04,kom04,kom05,dhk03,dv05a}.  Unlike in the radial direction
through the disk \citep{st90,igumenshchev2003,narayan2003}, the jet is
interchange {\it stable} across the interface between the disk wind
and the Poynting-dominated jet \citep{le93}, so these nonaxisymmetric
modes do not lead to significant mass-loading of the Poynting jet.
Anomalous resistivity due to reconnection between the wind and funnel
is not strong enough to significantly fill the funnel even at modest
resolutions that under-resolve the funnel-wall interface
\citep{mg04,mckinney2004}. Also, the disk wind, which straddles the
Poynting jet, has a sufficiently large angular momentum to avoid
significantly contaminating the Poynting jet (see, e.g.,
\citealt{moch93}). That is, unlike in the radial direction in the
disk, there is no angular momentum transport across the $\theta$
direction between the Poynting-dominated and the disk wind.
Therefore, to evolve the GRMHD equations of motion, which require
matter, a numerical scheme or physical mechanism must be invoked to
fill the funnel with matter once it evacuates.

As done by all others, in this paper the rest-mass and internal energy
density are limited to a small but finite value in the
Poynting-dominated jet by employing a so-called ``floor model.''  This
model forces a minimum on the rest-mass density ($\rho_{fl}$) and
internal energy density ($u_{fl}$), which are usually set to several
orders of magnitude lower than the disk density.  This effectively
modifies the equations of motion.

Floor-models necessarily violate the ideal MHD approximation because
they violate rest-mass and energy-momentum conservation.  This occurs,
at least, where the poloidal velocity $u^p=0$ at the stagnation
surface in the Poynting-dominated jet.  Matter inside the stagnation
surface falls into the black hole, while matter outside it is ejected
as part of the jet.  Arbitrary floor-models violate the ideal-MHD
condition far away from the black hole where no rest-mass should be
added for any physical mass-loading model
\citep{phi83,mckinney2004,mckinney2005b}. For example, if a numerical
model uses $\rho_{fl}\sim{\rm Const.}$ and the floor value is reached
at small radii, then the ideal-MHD approximation is violated for the
entire length of the jet in the funnel because the density would
otherwise have decreased with radius.

We use a floor model with a minimum allowed rest-mass density of
$\rho_{fl}=10^{-7} r^{-2.7}\rho_{0,disk}$ and minimum allowed internal
energy density of $u_{fl}=10^{-9} r^{-2.7}\rho_{0,disk}$. The
coefficients are not chosen arbitrarily small or large for two
reasons.  First, the value of $b^2/\rho_0$ in the jet determines the
magnetic energy per unit particle, which determines the maximum
Lorentz factor the jet can obtain.  As long as $b^2/\rho_0>1$ near the
poles, then we find that a relativistic jet emerges around the poles.
The important physics in the emergence of a BZ-driven jet is that the
gradient in the toroidal magnetic field dominates the gas+ram pressure
of the surrounding material.  The density floor is chosen to allow the
emergence of a fast jet since the magnetic energy per unit particle is
large ($b^2/\rho_0\lesssim 10^7$ since $b^2\sim \rho_{0,disk} c^2$ is
expected near the poles of the black hole) in the jet once it is
evacuated.  Conversely, if a density floor coefficient is chosen such
that $\rho_{fl}\gtrsim 10^{-1}\rho_{0,disk}$, then we find that no
Poynting-dominated jet emerges since the value of $b^2/\rho_0\lesssim
1$ around the black hole near the poles.  A self-consistent model of
the mass-loading must at least determine the time-averaged density
near the black hole around the poles. Second, the floor coefficient is
chosen to be not too low in order to avoid numerical difficulties in
integrating the equations when the floor is activated in regions of
large magnetic energy density per unit rest-mass density.  This is a
common problem with all present GRMHD numerical methods since the
equations of motion become a stiff set of equations in the limit
$b^2/\rho_0\gg 1$.

The radial scaling of the floor model is chosen to avoid contamination
in the jet by the floor model at large radii.  At large radii the
rest-mass density drops off $\propto r^{-2.2}$, so the rest-mass
density scaling of $\propto r^{-2.7}$ does not interfere with this.  A
radial scaling with a power $\lesssim 2.2$ would lead to nonphysical
results by loading the jet with extra high-velocity material.  The
internal energy density floor is never reached in the jet due to the
shock heating, which keeps $u/\rho_0>1$ in the jet.  See
section~\ref{results} for a more detailed discussion.

Our code keeps track of how often the density goes below the floors
and how this modifies the conservation of mass, energy, and angular
momentum.  The floor model contribution is negligible except in the
Poynting-dominated jet within $r\lesssim 10r_g$.  This injection of
mass is a simple ad-hoc model for the mass-loading of the Poynting
jet.  Beyond $r\approx 10r_g$, the floor is not activated because the
accumulated rest-mass is larger than the floor value.  Thus, beyond
this point the jet is self-consistently evolved strictly within the
ideal MHD approximation.

No explicit reconnection model is included.  However, our code checks
the effective resistivity by measuring the rest-mass flux across field
lines.  For the boundary between the disk wind and the
Poynting-dominated jet, the total rest-mass flux across the field line
is negligible compared to the rest-mass flux along the field line
determined by the floor model.  An unresolved model would load field
lines with rest-mass that crosses field lines and not properly
represent any physical model of resistivity \citep{mckinney2004}.

This ``floor model'' prescription was found to give quantitatively
similar results to physically-based models of the mass-loading
\citep{mckinney2005c} for which most of the mass-loading takes place
only near the black hole for $r\lesssim 10r_g$.  Note that
\citet{mckinney2005c} performed identical (including random seed for
initial condition perturbations) simulations as in this paper except
the method used to add mass and energy to the Poynting jet.  In this
paper, the floor model is chosen to most closely match the
physically-based model described in \citet{mckinney2005c}.  As
mentioned there, the results are quantitatively similar except within
$r<10r_g$ and only inside the Poynting jet.

\subsection{Lorentz Factors}\label{lorentzfactors}

This paper studies the speed of the jet as determined by the Lorentz
factor.  The Lorentz factor of the jet can be measured with respect to
any frame.  We measure the Lorentz factor with respect to a static
observer at infinity,
\begin{equation}
\Gamma\equiv u^{\hat{t}}=u^t\sqrt{-g_{tt}} ,
\end{equation}
in Boyer-Lindquist coordinates, where no static observers exist inside
the ergosphere.  This is in contrast to the definition of $W\equiv
u^t\sqrt{-1/g^{tt}}$, which is the Lorentz factor as measured by a
ZAMO observer as used by many numerical relativists.

Strict upper limits on the Lorentz factor and the angular velocity at
large distances can be obtained from the local flow quantities.  The
Lorentz factor and $\phi$-velocity at large distances for an
axisymmetric, stationary flow have upper limits of
\begin{eqnarray}
\Gamma_{\infty}         & \equiv & \frac{-T^p_t}{\rho_0 u^p} = -h u_t    + \Phi\Omega_F B_\phi  \\
u^{\hat{\phi}}_{\infty} & \equiv & \frac{T^p_\phi}{\rho_0 u^p} = h
u_\phi + \Phi B_\phi ,
\end{eqnarray}
where $p$ indicates either poloidal direction (e.g., $r$ or $\theta$),
$h\equiv(\rho_0+\IEDEN+p)/\rho_0$ is the specific enthalpy,
$\Phi\equiv B^p/(\rho_0 u^p)$ is the conserved magnetic flux per unit
rest-mass flux, $\Omega_F\equiv F_{tp}/F_{p\phi}$ is the conserved
field rotation frequency, $B_\phi\equiv \dF_{\phi t}$ is the covariant
toroidal magnetic field, and $u^{\hat{\phi}}_{\infty}\equiv
u_{\phi,\infty}$ is the orthonormal angular velocity at large
distances.  For a stationary, axisymmetric flow, these are the energy
and angular momentum flux per unit rest-mass flux that are conserved
along flux surfaces.  The matter and electromagnetic pieces are
separable, such that
\begin{eqnarray}
\Gamma_{\infty} & = & \Gamma^{{\rm (MA)}}_\infty + \Gamma^{{\rm (EM)}}_\infty  \\
u_{\phi,\infty} & = & u^{{\rm (MA)}}_{\phi,\infty} + u^{{\rm {\rm (EM)}}}_{\phi,\infty} .
\end{eqnarray}
See \citet{mckinney2005b} for more details.

Excluding radiative processes and with some physical accounting of the
mass-loading of the jet, the results of this paper suggest one can
approximately assume almost all the magnetic and thermal energy go
into particle kinetic energy.  Thus $\Gamma_\infty$ and
$u_{\phi,\infty}$ may represent the Lorentz factor and angular
velocity at large distances before the energy is lost to radiation
once the flow is optically thin.

\subsection{Units}\label{units}

The units in this paper have $G M = c = 1$, which sets the scale of
length ($r_g\equiv GM/c^2$) and time ($t_g\equiv GM/c^3$).  The mass
scale is determined by setting the observed (model-dependent measured
or inferred for GRB-type systems) mass accretion rate ($\dot{M}_0$)
equal to the accretion rate through the black hole horizon as measured
in a simulation.  So the mass scale is scaled by the mass accretion
rate ($\dot{M}_0$) at the horizon ($r=r_H\equiv r_g(1+\sqrt{1-j^2})$),
such that $\rho_{0,disk}\equiv \dot{M}_0[r=r_H] t_g/r_g^3$ and the
mass scale is then just $m\equiv \rho_{0,disk} r_g^3 =
\dot{M}_0[r=r_H] t_g$.  For a black hole with angular momentum $J=j
GM^2/c$, $j=a/M$ is the dimensionless Kerr parameter with $-1\le j\le
1$.

The results of the simulations can be applied to any astrophysical
system once the value of $\rho_{0,disk}$ is estimated.  For example, a
collapsar model with $\dot{M}=0.1\msun s^{-1}$ and $M\approx 3\msun$,
then $\rho_{0,disk}\approx 3.4\times 10^{10}{\rm g}\cm^{-3}$
\citep{mw99}.  M87 has a mass accretion rate of $\dot{M}\sim
10^{-2}\msun\yr^{-1}$ and a black hole mass of $M\approx 3\times
10^9\msun$ \citep{ho99,reynolds96} giving $\rho_{0,disk}\sim 10^{-16}
{\rm g}\cm^{-3}$.  GRS 1915+105 has a mass accretion rate of
$\dot{M}\sim 7\times 10^{-7}\msun\yr^{-1}$ \citep{mr94,mr99,fb04} with
a mass of $M\sim 14\msun$ \citep{greiner2001a}, but see
\citet{kaiser04}.  This gives $\rho_{0,disk}\sim 3\times 10^{-4}{\rm
g}\cm^{-3}$.

\section{Numerical Setup}\label{setup}

The GRMHD equations of motion are integrated numerically using a
modified version of a numerical code called HARM \citep{gmt03}, which
uses a conservative, shock-capturing scheme. Compared to the original
HARM, the inversion of conserved quantities to primitive variables is
performed by using a new faster and more robust two-dimensional
non-linear solver \citep{noble05}.  A parabolic interpolation scheme
\citep{colella84} is used rather than a linear interpolation scheme,
and an optimal total variational diminishing (TVD) third order
Runge-Kutta time stepping \citep{shu97} is used rather than the
mid-point method.  For the problems under consideration, the parabolic
interpolation and third order time stepping method reduce the
truncation error significantly, including regions where $b^2/\rho_0\gg
1$.

\subsection{Computational Domain}

The computational domain is {\it axisymmetric}, with a grid that
extends from $r_{in} = 0.98 r_H$, where $r_H$ is the horizon, to
$r_{out} = 10^4r_g$, and from $\theta = 0$ to $\theta = \pi$.  Our
numerical integrations are carried out on a uniform grid with
coordinates: $x_0, x_1, x_2, x_3$, where $x_0 = t[{\rm Kerr-Schild}]$,
$x_3 = \phi[{\rm Kerr-Schild}]$.  The radial coordinate is chosen to
be
\begin{equation}\label{radius}
r = R_0 + e^{x^{n_r}_1} ,
\end{equation}
where $R_0$ is chosen to concentrate the grid zones toward the event
horizon (as $R_0$ is increased from $0$ to $r_H$) and $n_r$ is
controls the enhancement of inner to outer-radial regions.  For
studies where the disk and jet interaction is of primary interest,
$R_0=0$ and $n_r=1$ are chosen.  For this paper, for which in addition
the far-field jet is of interest, $R_0=-3$ and $n_r=10$ are chosen in
order to reach large distances with few grid points and to have the
inner-radial computational region resolved similarly to the models of
\citet{mg04}. The $\theta$ coordinate is chosen to be
\begin{equation}\label{theta}
\theta = \pi x_2 + \frac{1}{2} (1 - h(r)) \sin(2\pi x_2) ,
\end{equation}
where $h(r)$ is used to concentrate grid zones toward the equator (as
$h$ is decreased from $1$ to $0$) or pole (as $h$ is increased from
$1$ to $2$).  The jet at large radii is resolved together with the
disk at small radii using
\begin{equation}\label{hofr}
h(r) = 2 - Q_j (r/r_{0j})^{-n_j g_j}
\end{equation}
with the parameters of $Q_j=1.3$, $r_{0j}=2.8$, $n_j=0.3$,
$r_{1j}=20$, $r_{2j}=80$, and
\begin{equation}
g_j=g_j(r)=\frac{1}{2}+\frac{1}{\pi}{\rm atan}(\frac{r-r_{2j}}{r_{1j}})
\end{equation}
is used to control the transition between inner and outer-radial
regions.  The values of $Q_j$, $r_{0j}$, $n_j$, $r_{1j}$, and $r_{2j}$
were set empirically to approximately follow the simulated jet from
the horizon to large distances.  An alternative to this fixed
refinement of the jet and disk is an adaptive refinement (see, e.g.,
\citealt{zm05}).

The numerical resolution of all the models described is $512\times
256$ compared to the fiducial model with $456^2$ in \citet{mg04}.
However, due to the enhanced $\theta$ grid, the resolution in the
far-field jet region is $\sim 10$ times larger.  Also, with the use of
a parabolic interpolation scheme, the overall effective resolution is
additionally enhanced.  Compared to our previous model this gives us
an effective $\theta$ resolution of $\approx 9000$ zones.

All the results in this paper are robust to changes in numerical
resolution, as has been explicitly verified by convergence testing in
which the resolution, order of the spatial interpolation, and
$\theta$-grid geometry were varied.  This included varying the
transition radius at which the grid resolution in the jet becomes
larger than the disk via different $Q_j=\{1.3,1.8\}$, using
resolutions of $256\times 256$ and $512\times 256$, and using linear
and parabolic spatial interpolation schemes.  In all cases, no
qualitative differences were found.

\subsection{Model Setup}

All the experiments performed evolve an initially weakly magnetized
torus around a Kerr black hole in axisymmetry.  A generic model is
used so the results mostly can be applied to any black hole system.
The model of the initial disk is identical to the model studied in
\citet{mg04}, and so direct comparisons can be made to our prior work.
The new models have an extended radial grid and have a longer
duration.  These were required to study the large-scale
propagation of the Poynting-dominated jet.

A black hole spin of $j=0.9375$ is chosen, but this produces similar
results to models with $0.5\lesssim j\lesssim 0.99$ (see also, e.g.,
\citealt{mg04}).  This value of the black hole spin is close to the
equilibrium black hole spin for a black hole accreting a disk with a
height ($H$) to radius ($R$) ratio of $H/R\sim 0.26$ \citep{gsm04}.

The model is run for $\Delta t = 1.4\times 10^4t_g$, which is about
$52$ orbital periods at the pressure maximum and about $1150$ orbital
periods at the black hole horizon.

Since the model is axisymmetric, disk turbulence is not sustained
after about $t\sim 3000t_g$ \citep{mg04}, after which the anti-dynamo
theorem prevails \citep{cowling34}.  However, while this affects the
disk accretion by forcing it to become more ``laminar,'' this does not
affect the evolution of the Poynting-dominated jet.  That is, from the
time of turbulent accretion to ``laminar'' accretion, the funnel
region is mostly unchanged.  Indeed, the far-field jet that has
already formed is causally disconnected from the region where the
accretion disk would still be turbulent.

\subsection{Initial Conditions}\label{ic}

The initial conditions consist of an equilibrium torus, which is a
``donut'' of plasma with a black hole at the center
\citep{fm76,ajs78}.  This torus is a solution to the axisymmetric
stationary hydrodynamic equations and is supported against gravity by
centrifugal and pressure forces. The solutions of \cite{fm76}
corresponding to $u^t u_\phi = const.$ are used. The initial inner
edge of the torus is set at $r_{edge} = 6$. The equation of state is a
gamma-law ideal gas with $\gamma = 4/3$, but other $\gamma$ lead to
similar results \citep{mg04}.  The models have $u^t u_\phi = 4.281$,
the pressure maximum is located at $r_{max} = 12r_g$, the inner edge
at $(r,\theta)=(6r_g,\pi/2)$, and the outer edge at $(r,\theta) =
(42r_g, \pi/2)$.  The orbital period at the pressure maximum is $2 \pi
(a + \left(r_{max}/r_g\right)^{3/2}) t_g \simeq 267 t_g$, as measured
by an observer at infinity.  The torus chosen has a disk height ($H$)
to radius ($R$) ratio of $H/R\sim 0.26$ on average, and is slightly
thinner (thicker) at the inner (outer) radial edge.

As done by \citet{mg04}, a purely poloidal magnetic field is placed
into the initial torus. The field can be described using a vector
potential with a single nonzero component, and we choose $A_\phi
\propto {\rm MAX}(\rho_0/\rho_{0,max} - 0.2, 0)$. The field is
therefore restricted to regions with $\rho_0/\rho_{0,max} > 0.2$.  The
restriction of the field to be inside the torus is to avoid regions
that are numerically estimated to have large magnetic field per unit
particle just outside the torus.  The field is normalized so that the
minimum ratio of gas to magnetic pressure is $100$.  The equilibrium
is therefore only weakly perturbed by the magnetic field.

While the torus is stable to axisymmetric hydrodynamic instabilities,
the torus is unstable to global nonaxisymmetric hydrodynamic modes
\citep{pp83}.  However, when the torus is embedded in a weak magnetic
field, the magnetorotational instability (MRI) dominates those
hydrodynamic modes \citep{bh91,gam04}. Small perturbations are
introduced in the internal energy density, which induces velocity
perturbations that seed the instability.

No initial large-scale net vertical field is necessary to drive a
Poynting jet, since a large-scale poloidal field is self-consistently
generated with the above prescription for the initial field geometry
\citep{hirose04,mg04,mckinney2005a}.  In addition, other field
geometries, such as a net vertical field or multiple field loops
within the disk, lead to similar results \citep{mg04}.  Reconnection
efficiently erases the initial geometrical differences
\citep{mg04,mckinney2005a}.  Only contrived fields without any
poloidal component lead to no Poynting jet \citep{dv05a}.  Otherwise,
the magneto-rotational instability drives the amplification of
toroidal and poloidal flux through differential rotation.

Beyond the disk, we choose two different models of the matter.  The
first is to set the densities to the ``floor model'' value (denoted as
environment 1).  The second is to set the rest-mass density to $\rho_0
= 10^{-4} r^{-3/2}\rho_{0,disk}$ and the internal energy density to $u
= 10^{-6} r^{-5/2}\rho_{0,disk}$ (denoted as environment 2).  In both
cases, there is no magnetic field outside the disk.  The velocity
everywhere is set to be equal to the free-falling frame.  The fiducial
model described below focuses on the second model of the surrounding
medium, while the first model is used to demonstrate the effect of the
existence of an environment on the jet.

\subsection{Boundary Conditions}\label{bc}

Two different models are chosen for the outer boundary condition.  The
first model is to use a so-called ``outflow'' boundary condition, for
which all primitive variables ($\{\rho_0,u,\tilde{u}^i,B^i\}$) are
projected into the ghost zones while forbidding inflow (associated
with environment 1).  The second model is to inject matter at some
specified rate at the outer boundary, unless there is outflow.  For
the second model, unless there is outflow the outer boundary is set to
inject mass at the free-fall rate with no angular momentum, with a
density of $\rho_0 = 10^{-4} r^{-3/2}\rho_{0,disk}$ and $u = 10^{-6}
r^{-5/2}\rho_{0,disk}$ (associated with environment 2).

The fiducial model below uses the second model, while the first model
is used together with the ``floor model'' version of the initial
conditions in order to investigate the effect of the environment on
the jet.

\subsection{Astrophysical Applicability of Model}\label{appmodel}

As applied to the collapsar model, the outer grid radius of $r_{out} =
10^4r_g$ corresponds to about $20$ presupernova core radii
\citep{ww95}, $\sim 10^{10}\km$, or about $1/10$th the entire star's
radius.  As applied to M87, the outer radius of the grid is at
$1.4\pc$.  For x-ray binaries, the outer radius of the grid is at
$0.001$AU.

For the collapsar model, the chosen black hole spin would be achieved
in the late phase of jet production \citep{mw99}.  As applied to AGN
and x-ray binaries, the results should be approximately applicable to
systems with $0.5\lesssim j\lesssim 0.99$.

The duration of the evolution for the collapsar model is $\sim 0.2$
seconds, and at the spin chosen would correspond to the late phase of
accretion when the Poynting flux reaches its largest magnitude
\citep{mw99}.  For the AGN M87 the duration corresponds to $\sim
7~{\rm years}$.  For the x-ray binary GRS 1915+105 the duration
corresponds to $\sim 1~{\rm second}$.

The chosen $H/R$ and the extent of the disk are similar to the
accretion disk that is expected to form in the collapsar model
\citep{mw99,kohri2002,kohri2005}.  While in AGN and x-ray binary
systems an extended disk should be studied, the extent of the disk
does not affect the results of the jet produced since the time scale
of the simulations are relatively short compared to the outer disks'
accretion time scale, $t\sim 2\pi/(\alpha \Omega_K)\sim 10^5t_g$,
where $\alpha\sim 0.01$ and $\Omega_K$ is the Keplerian angular
frequency.  The disk thickness is similar to the disk thickness of
radiatively inefficient disk models that are used to represent the
disks in some AGN and x-ray binaries during their radiatively
inefficient cycle \citep{ny95}. A study of the effect of disk
thickness is left for future work.

The ``Bondi-flow'' model of the matter beyond the disk (environment 2)
can be compared to the collapsar model. This would correspond to the
collapsing envelope of a massive star.  Presupernova models suggest
that the infalling matter is at about $30\%$ the free-fall rate we
have chosen \citep{mw99}, but this small percent difference is
unlikely to significantly affect the jet formation.  In addition, the
angular momentum of the envelope is not important since the disk is
already present in our model and the timescale for adding mass to the
disk in an accretion shock is much longer than the duration of the
simulation.  In collapsar models, the density structure near the
equatorial plane varies between $\rho_0\propto r^{-3/2}$ to
$\rho_0\propto r^{-2}$ and the internal energy is $u\propto r^{-5/2}$
to $u\propto r^{-2.7}$ \citep{mw99}.  This is similar to our model,
which is expected since the collapsar model involves mostly spherical
collapse.  As applied to AGN and x-ray binaries, the second model
(environment 2) is a type of Bondi inflow of the surrounding medium.

The fiducial model below uses the second model, while the first model
is used together with the ``floor model'' version of the initial
conditions in order to investigate the effect of the environment on
the jet.

\section{Numerical Results}\label{results}

The overall character of the accretion flow is unchanged compared to
the descriptions given in \citet{mg04}.  The disk enters a long,
quasi-steady phase in which the accretion rates of rest-mass, angular
momentum, and energy onto the black hole fluctuate around a
well-defined mean.  Beyond $t\sim 3000t_g$, the mass accretion rate
drops, but all other quantities in the disk and jet remain similar.

As in \citet{mg04}, a mildly relativistic wind is launched from the
inner edge of the disk with a half-opening angle of roughly $16^\circ$
to $45^\circ$.  In this model the disk wind has $\Gamma\sim 1.5$.
However, a long temporal study of this disk wind depends on sustaining
disk turbulence, which decays beyond $t\sim 3000t_g$ in this
axisymmetric model.

The coronal-funnel boundary contains shocks with a sonic Mach number
of $M_s\sim 100$.  The inner-radial interface between the disk and
corona is a site of vigorous reconnection due to the magnetic buoyancy
and convective instabilities present there.  These two parts of the
corona (coronal-funnel wall and inner-radial disk-corona interface)
are about $100$ times hotter than the bulk of the disk.  Thus, these
coronal components are a possible sites for Comptonization and
nonthermal particle acceleration.

Meanwhile, as in \citet{mg04}, a Poynting-dominated jet has formed.
The Poynting-dominated jet forms as the differential rotation of the
disk and the frame-dragging of the black hole induce a significant
toroidal field that launches material away from the black hole.  The
Poynting-dominated jet is launched once the pressure of the mostly
nonmagnetic initial funnel material is lower than the toroidal
magnetic pressure.  This occurs within $t\lesssim 500t_g$.

\begin{figure}
\subfigure{\includegraphics[width=3.3in,clip]{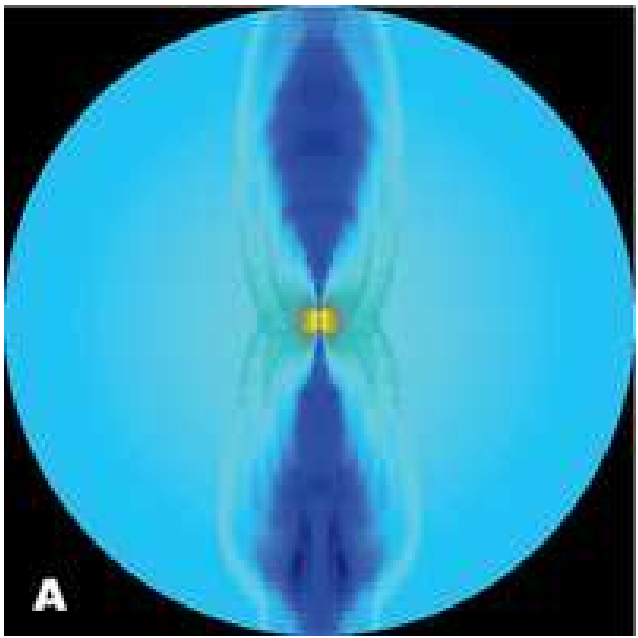}}
\subfigure{\includegraphics[width=3.3in,clip]{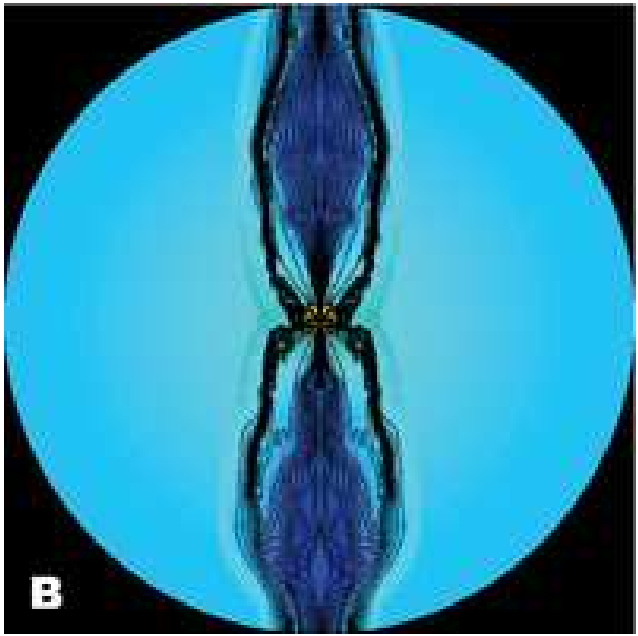}}

\caption{Panel (A) shows final distribution of $\log{\rho_0}$ on the
Cartesian plane. Black hole is located at center. Panel (B) shows
magnetic field overlaid on top of log of density. Outer scale is
$r=10^4 r_g$. Time is $t=1.4\times 10^4t_g$.  Color represents
$\log(\rho_0/\rho_{0,disk})$ with dark red highest and dark blue
lowest.  The final state has a density maximum of $\rho_0\approx
2\rho_{0,disk}$ and a minimum of $\rho_0\sim 10^{-13}\rho_{0,disk}$ at
large radii.  Grid zones are not smoothed to show grid structure.
Outer-radial zones are large, but outer $\theta$ zones are below the
resolution of the figure.  For the purposes of properly visualizing
the accretion flow and jet, we follow \citet{mw99} and show both the
negative and positive $x$-region by duplicating the axisymmetric
result across the vertical axis.}
\label{density}
\end{figure}

\begin{figure}
\subfigure{\includegraphics[width=3.3in,clip]{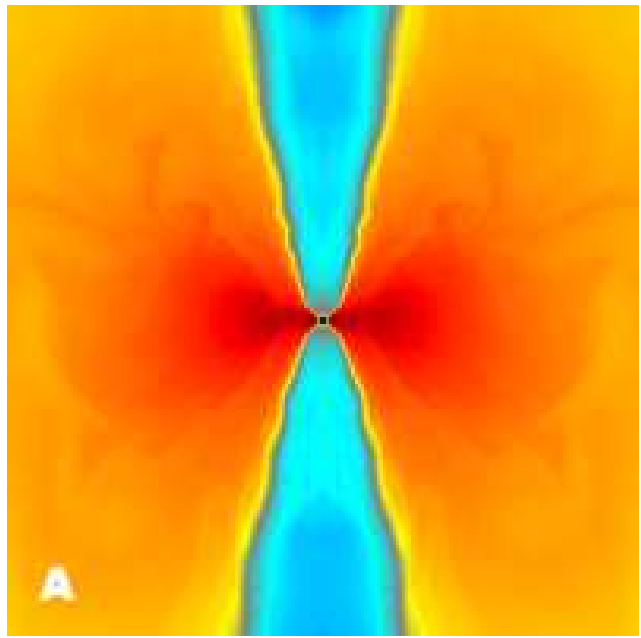}}
\subfigure{\includegraphics[width=3.3in,clip]{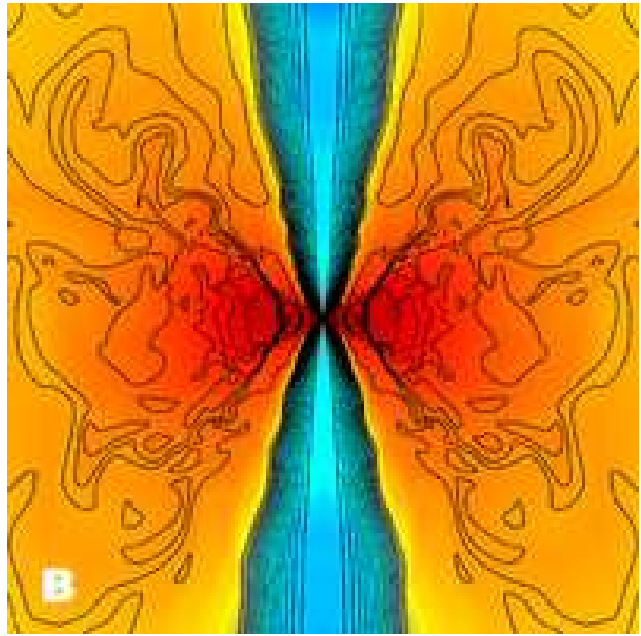}}

\caption{Same as figure~\ref{density}, but outer scale is $r=10^2 r_g$. }

\label{densityzoom}
\end{figure}

Figure~\ref{density} and figure~\ref{densityzoom} show the log of
rest-mass density and magnetic field at the final time of $t=1.4\times
10^4t_g$.  Clearly, the jet has pummelled its way through the
surrounding medium.  By the end of the simulation, the field has been
self-consistently launched into the funnel region and has a regular
geometry there.  The density and field show how the surrounding medium
is shocked by the jet in a lateral direction. In the disk and at the
surface of the disk the field is curved on the scale of the disk scale
height.  Within $r\lesssim 10^2r_g$ the funnel field is ordered and
stable due to the poloidal field dominance.  However, beyond $r\sim
10-10^2r_g$ the poloidal field is relatively weak compared to the
toroidal field and the field lines bend and oscillate erratically due
to pinch instabilities.  The radial scale of the oscillations is $10^2
r_g$ (but up to $10^3r_g$ and as small as $10r_g$), where $r\sim
10r_g$ is the radius where poloidal and toroidal field strengths are
equal.  By the end of the simulation, the jet has only fully evolved
to a state independent of the initial conditions out to $r\approx
5\times 10^3 r_g$, beyond which the jet features are a result of the
tail-end of the initial launch of the field.  The head of the jet has
passed beyond the outer boundary of $r=10^4 r_g$.

Figure~\ref{densityzoom} shows that the strongest magnetic field is
near the black hole in an X-configuration due to the Blandford-Znajek
effect having a power $dP_{jet}/d\theta \propto \sin^2{\theta}$ that
is truncated by the disk near the equatorial plane.

\begin{figure}
\includegraphics[width=3.3in,clip]{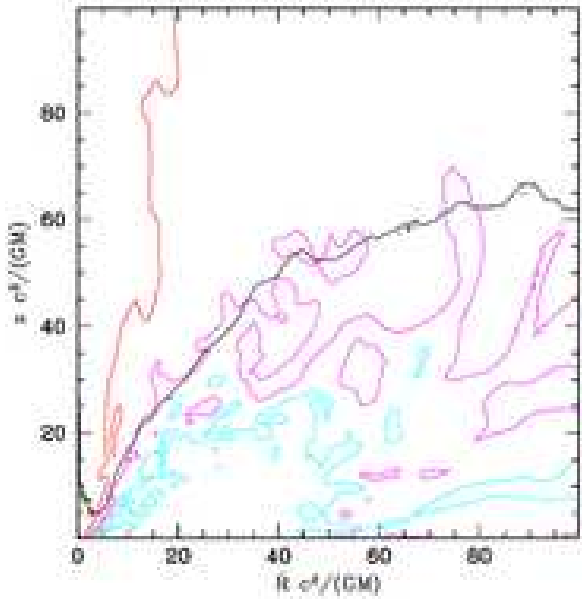}

\caption{ Contours for the disk-corona, corona-wind, and wind-jet
boundaries at $t\approx 1500t_g$ and an outer scale of $r\approx
10^2r_g$.  The disk-corona boundary is a cyan contour where
$\beta\approx u/b^2=3$, the corona-wind boundary is a magenta contour
where $\beta=1$, and the wind-jet boundary is a red contour where
$b^2/(\rho_0 c^2)=1$.  The black contour denotes the boundary beyond
which material is unbound and outbound (wind + jet). }
\label{linesi}
\end{figure}

\begin{figure}
\includegraphics[width=3.3in,clip]{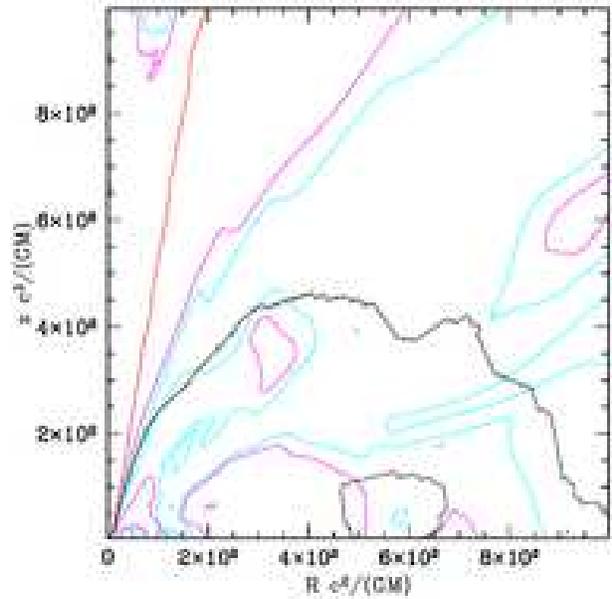}

\caption{Same as figure~\ref{linesi}, but for $t\approx 1.4\times
10^4t_g$ and an outer scale of $r\approx 10^3r_g$.  At large scales,
the cyan and magenta contours closer to the equatorial plane are not
expected to cleanly distinguish any particular structure. }
\label{lineso}
\end{figure}

\begin{figure}
\includegraphics[width=3.3in,clip]{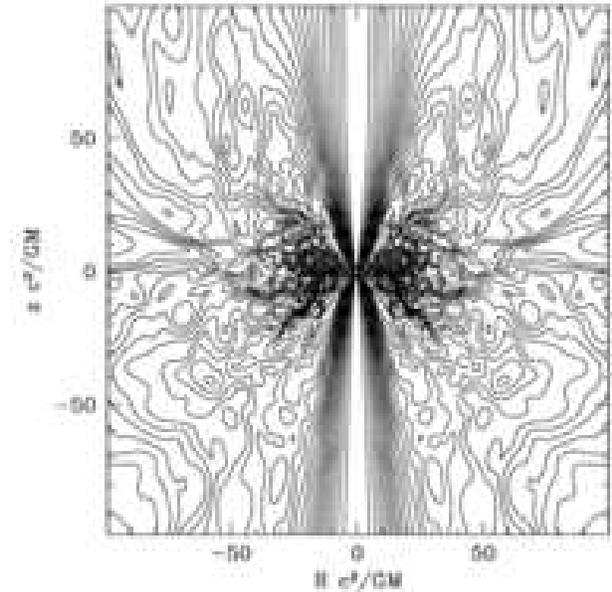}

\caption{Field geometry near black hole for $t\approx 1500t_g$ during
  phase of strong disk turbulence. }
\label{iaphizoom}
\end{figure}

Figures~\ref{linesi} and~\ref{lineso} show the energy structure of
the disk, corona, and jet.  This figure is comparable to the left
panel of figure 2 in \citet{mg04}.  These plots show the boundaries
between the ``disk,'' ``corona,'' ``disk wind,'' and ``Poynting
jet.''  The disk-corona boundary is defined to be where $\beta\equiv
p_g/p_b \equiv 2(\gamma-1)u/b^2 = 3$, the corona-wind boundary is
defined to be where $\beta=1$, and the wind-jet boundary is defined
to be where $b^2/(\rho_0 c^2)=1$.  The plot also shows the boundary
beyond which material is unbound and outbound (wind and jet), which
here is defined as where $u^p>0$ and $u_t=-1$.  This corresponds to
where non-interacting {\it particles} are unbound, although thermal
and magnetic energy can further unbind the matter.  Hence, more
material within the surface is also unbound.

The jet stays collimated, while the disk wind has a large opening
angle at large radii. Beyond $r\approx 10r_g$, the jet undergoes
poloidal oscillations due to toroidal pinch instabilities. By
$r\gtrsim 100r_g$, pinch instabilities subside once magnetic energy is
converted into thermal energy, which then supports the jet. Residual
slow dense and fast diffuse patches from the instability are present
in the jet, such as the slower and cooler dense blob shown at top left
corner of figure.

Figure~\ref{iaphizoom} shows the disk, corona, and jet magnetic field
structure during the turbulent phase of accretion at $t\approx
1500t_g$.  Compared to figure~\ref{densityzoom}, this shows the
turbulence in the disk, but is otherwise similar.  The jet, disk, and
coronal structures remain mostly unchanged at late times despite the
decay of disk turbulence.  That is, the current sheets in the disk do
not decay and continue to support the field around the black hole that
leads to the Blandford-Znajek effect.  The coronal thickness and
radial extent do not require the disk turbulence, which only adds to
the time-dependence of the disk wind.  Note that despite the fact that
the disk wind only contains small-scale magnetic fields, such fields
still lead to self-collimation \citep{li2002}.

\begin{figure}
\subfigure{\includegraphics[width=3.3in,clip]{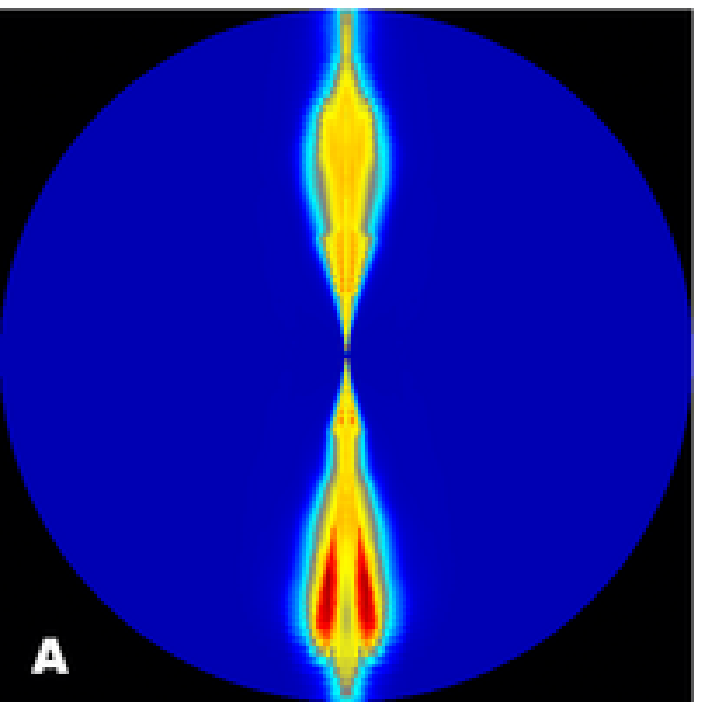}}
\subfigure{\includegraphics[width=3.3in,clip]{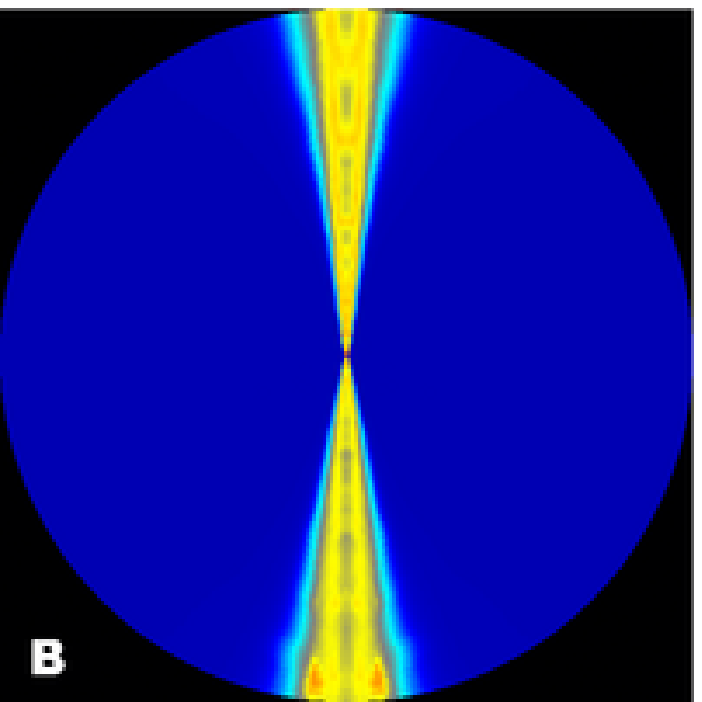}}

\caption{Plot of $\log{\Gamma_\infty}$ with red highest
($\Gamma_\infty\sim 10^4$) and blue lowest.  Yellow is
$\Gamma_\infty\sim 10^2 - 10^3$. Panel (A) shows outer scale of
$r=10^4 r_g$.  Panel (B) shows outer scale of $r=10^3 r_g$ with same
color scale.  Jet is independent of initial conditions by $r\approx
5\times 10^3r_g$. Jet becomes conical at large radii with a core
half-opening angle $\theta_j\approx 5^\circ$. }
\label{gammainf}
\end{figure}

Figure~\ref{gammainf} shows $\Gamma_\infty$, which reaches up to
$\Gamma_\infty \sim 10^3 - 10^4$, for an outer scale of $r=10^4 r_g$
(panel A) and an outer scale of $r=10^3r_g$ (panel B).  The
inner-radial region is not shown since $\Gamma_\infty$ is divergent
near the injection region where the ideal-MHD approximation breaks
down.  Different realizations (random seed of perturbations in disk)
lead up to about $\Gamma_\infty\sim 10^4$ as shown for the lower pole
in the color figures.  This particular model was chosen for
presentation for its diversity between the two polar axes.  The upper
polar axis is fairly well structured, while the lower polar axis has
undergone an atypically strong magnetic pinch instability.  Various
realizations show that the upper polar axis behavior is typical, so
this is studied in detail below.  The strong hollow-cone structure of
the lower jet is due to the strongest field being located at the
interface between the jet and the surrounding medium, and this is
related to the fact that the BZ-flux is $\propto \sin^2{\theta}$,
which vanishes identically along the polar axis.  It is only the
disk+corona that has truncated the energy extracted, otherwise the
peak power would be at the equator.  As described in the next section,
at larger radii the jet undergoes collimation that results in more of
the energy near the polar axis.

\subsection{Radial Jet Structure}\label{radialstructure}

\begin{figure}

\includegraphics[width=3.3in,clip]{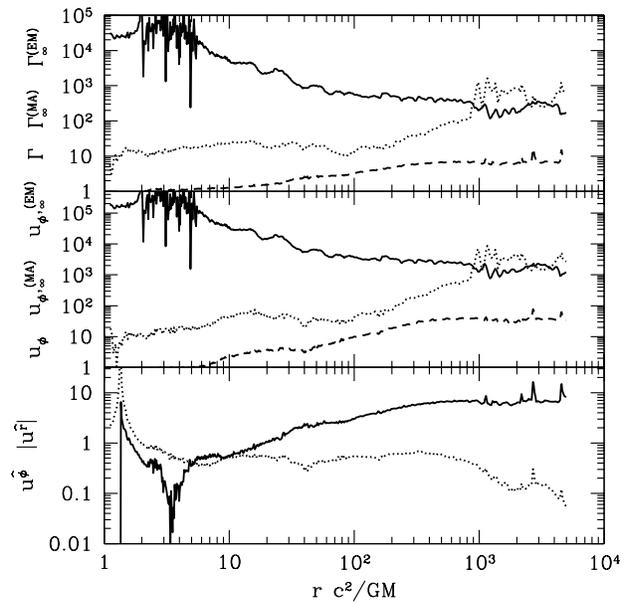}
\caption{A radial cross section along a mid-level field line in the
jet showing the velocity structure.  The top panel shows
$\Gamma^{{\rm (EM)}}_\infty$ (solid line), $\Gamma^{{\rm (MA)}}_\infty$ (dotted
line), and $\Gamma$ (dashed line).  The middle panel shows
$u_{\phi,\infty}^{{\rm (EM)}}$ (solid line), $u_{\phi,\infty}^{{\rm (MA)}}$
(dotted line), and $u_{\phi}$ (dashed line).  The bottom panel shows
$|u^{\hat{r}}|$ (solid line) and $u^{\hat{\phi}}$ (dotted line).}
  \label{gammas}
\end{figure}

Figure~\ref{gammas} shows the velocity structure of the
Poynting-dominated jet along a mid-level field line, which starts at
$\theta_j\approx 46^\circ$ on the black hole horizon and goes to large
distance, where the funnel region starts at $\theta_j\approx 63^\circ$
on the horizon.  The quantities shown are time-averaged values at
late-time.

The top panel of figure~\ref{gammas} shows the electromagnetic (EM)
energy flux per unit rest-mass flux ($\Gamma^{{\rm (EM)}}_\infty$),
the matter (MA) energy flux per unit rest-mass flux ($\Gamma^{{\rm
(MA)}}_\infty$), and the Lorentz factor as measured by an observer at
infinity ($\Gamma$).  For $2\lesssim r\lesssim 10$, the value of
$\Gamma^{{\rm (EM)}}_\infty$ is highly oscillatory.  This shows the
location of the stagnation surface, where the poloidal velocity
$u^p=0$, is temporally variable.  The stagnation surface is, at least,
where the ideal-MHD approximation breaks down and thus at any moment
the value of $\Gamma^{{\rm (EM)}}_\infty$ nearly diverges.  Notice
that while at $r\lesssim 10^3r_g$ the Poynting flux dominates, the
Poynting flux energy is slowly converted into enthalpy flux due to
nonlinear time-dependent shock heating.  Shocks are expected beyond
the magneto-fast surface (see, e.g., \citealt{bt05}) and here are
partially due to pinch instabilities
\citep{e93,begelman1998,sikora2005}.

For $r\gtrsim 10^3r_g$, the enthalpy flux and Poynting flux are in
equipartition ($\Gamma^{{\rm (MA)}}_\infty\sim \Gamma^{{\rm
(EM)}}_\infty$) and so an equipartition ``magnetic fireball'' has
formed.  The terminal Lorentz factor is of order $\Gamma_\infty\sim
400$ for this choice of flow line.  The Lorentz factor by $r\sim 10^4$
is still only $3\lesssim \Gamma\lesssim 15$, with smaller spatial
patches with $\Gamma\sim 25$.  For a radiation pressure-dominated and
optically thick flow, this implies there will be an extended
acceleration region where the magnetic fireball loses energy during
adiabatic expansion.

In a narrow region within the core of the jet (not shown in
figure~\ref{gammas}) within a half-opening angle of $\theta\approx
5^\circ$, the jet accelerates with
\begin{equation}\label{gammabulkcore}
\Gamma\sim \left(\frac{r}{10r_g}\right)^{0.3} ,~~~{\rm (inner)}
\end{equation}
for $10<r\lesssim 5\times 10^3r_g$, after which the initial conditions
dominate the flow.  Thus $\Gamma\sim 6$ has been reached in the core.
The middle and outer angular parts of the jet follow
\begin{equation}\label{gammabulkother}
\Gamma\sim \left(\frac{r}{3r_g}\right)^{0.45} ,~~~{\rm (inner)}
\end{equation}
for $3<r\lesssim 10^3r_g$, after which the field becomes
pseudo-conical and there is no sustained acceleration (or the flow
even decelerates).  Thus, $\Gamma\sim 10$ has been reached.  As
discussed below, the field is nearly paraboloidal for the radial range
over which there is such power-law acceleration.  The behavior of
$\Gamma(r)$ is in good agreement with an analytic study of
acceleration in a paraboloidal field that predicts $\Gamma\propto
r^{1/2}$ \citep{bn05}.  Thus our numerical result is consistent with
their analytical result.

The next lower panel of figure~\ref{gammas} shows that the
electromagnetic angular momentum is converted to enthalpy angular
momentum and that there is a conversion to particle angular momentum.

The bottom panel of figure~\ref{gammas} shows the approximate
orthonormal radial ($u^{\hat{r}}\approx \sqrt{g_{rr}} u^r$) and $\phi$
($u^{\hat{\phi}}\approx \sqrt{g_{\phi\phi}} u^\phi$) 4-velocities in
Boyer-Lindquist coordinates, where the approximation being made is
that the metric is dominated by the diagonal terms for $r\gtrsim
10r_g$.  The radially directed motion becomes relativistic, while the
$\phi$-component of the 4-velocity becomes sub-relativistic at large
distances.  However, the fluid and field rotation is still large
enough that it may stabilize the jet against $m=1$ kink instabilities,
as shown in the nonrelativistic case \citep{ocp03,nakamura2004} and
analytically in the relativistic case \citep{tom01}.

\begin{figure*}

\begin{center}
\includegraphics[width=6.6in,clip]{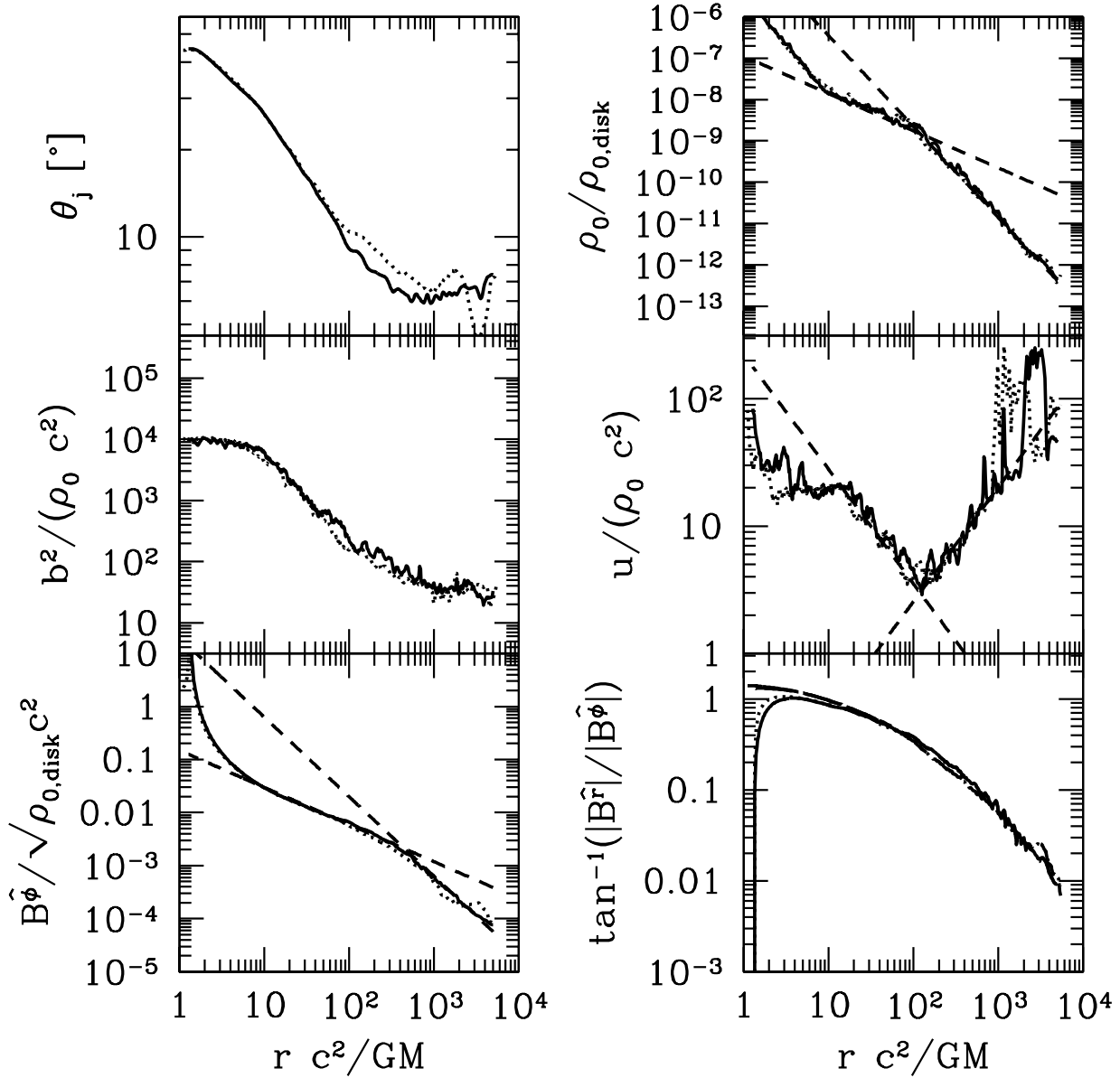}
\end{center}

\caption{ A radial cross-section of both poles (solid and dotted
lines) of the jet along a mid-level field line in the jet showing the
jet opening angle in degrees, density, and magnetic structure
($B^{\hat{r}}$ and $|B^{\hat{\phi}}|$ in Gaussian units). In the
density plots and the plot of $B^{\hat{\phi}}$, the overlapping dashed
lines are power-law fits. The pitch angle (lower right panel) in
Boyer-Lindquist coordinates (which vanishes at the horizon) has an
overlapping fit (dash-dotted line) to the \citet{tom01} kink stability
criteria when including field rotation. }
  \label{jetstruct2}
\end{figure*}

Figure~\ref{jetstruct2} shows the collimation angle, rest-mass and
internal energy densities, ratio of magnetic to rest-mass energy
density, toroidal field strength in orthonormal Boyer-Lindquist
coordinates, and the pitch angle along a mid-level field line in the
Poynting-dominated jet.  The field line is the same as that in
figure~\ref{gammas}, and now for both poles of the simulation.

The two poles are similar but not necessarily identical.  Many
simulations that were performed show at least one polar jet develops
some pinch instabilities.  In the model shown, one side develops very
strong instabilities by the end of the simulation.  The development of
the pinch instability mostly affects the far-radial Lorentz factor
($\Gamma$ and $\Gamma_\infty$ --- and so densities).  The value of
$\Gamma$ is up to 5 times larger in pinched regions than non-pinched
regions, while $\Gamma_\infty$ can be up to a factor 10 times larger
in pinched regions compared to non-pinched regions.

In figure~\ref{jetstruct2}, for $7r_g\lesssim r\lesssim 100r_g$ there
is a region of collimation slightly faster than power-law. For the
field line closest to the disk wind starting at $\theta_j\approx
57^\circ$, collimation is nearly a power-law with
\begin{equation}
\theta_j\approx 57^\circ \left(\frac{r}{2.8r_g}\right)^{-0.3}
\end{equation}
for $r<120r_g$ and $\theta_j\sim 8^\circ$ beyond.  Closer to the polar
axis the collimation is faster.  For the field line starting at
$\theta_j\approx 17^\circ$,
\begin{equation}
\theta_j\propto r^{-0.4} ,
\end{equation}
approximately up to $r\sim 120r_g$ and $\theta_j\sim 5^\circ$ beyond.
The inner core for $5\lesssim r\lesssim 10^3 r_g$ follows
$\theta_j\propto r^{-1/2}$, a paraboloidal field. See also discussion
related to figure~\ref{thetaj}.

From a measurement of the forces along and across the field lines
within the Poynting jet, the inner-radial collimation is dominated
by confinement by the disk+corona, while in the mid-radial range the
disk wind collimates the Poynting flow.  Far from the disk wind, the
internal jet is collimated by internal hoop stresses due to the
paraboloidal field.  Note that the classical hoop-stress paradigm
that jets can self-collimate is not fully tested here, but these
results suggest that collimation is in large part due to, or at
least requires, the disk and disk wind.

Notice from figure~\ref{gammas} that there is little acceleration
beyond $r\sim 10^3 r_g$.  It is well known that magnetic acceleration
requires field lines to collimate
\citep{e93,li1993,begli94,tomimatsu1994}.  In a conical flow, the
transfer of magnetic energy into kinetic energy is inefficient.  As
shown in figure~\ref{jetstruct2}, after $r\gtrsim 10^3 r_g$ the flow
oscillates around a pseudo-conical asymptote with a mean half-opening
angle of $\theta_j\sim 5^\circ$.  Beyond this radius, the magnetic
energy is instead converted to thermal energy.  The pseudo-conical
asymptote results once the magnetic energy no longer completely
dominates the other energy scales and the disk wind no longer helps
collimate the flow.  The final opening angle should be similar for
other black hole spins, as found in prior studies \citep{mg04}.

The lower panels of figure~\ref{jetstruct2} show the toroidal field
and pitch angle ($\alpha\equiv {\rm
tan}^{-1}(|B^{\hat{r}}|/|B^{\hat{\phi}}|)$), which shows that
eventually the toroidal field completely dominates the poloidal field,
and the toroidal field remains ordered. The toroidal field within
$r\lesssim 390r_g$ is well fit by
\begin{equation}
\frac{B^{\hat{\phi}}}{\sqrt{\rho_{0,disk}c^2}}[{\rm
G}]\approx -0.0023\left(\frac{r}{390r_g}\right)^{-0.7} .
\end{equation}
and is otherwise well fit by
\begin{equation}
\frac{B^{\hat{\phi}}}{\sqrt{\rho_{0,disk}c^2}}[{\rm
G}]\approx -0.0023\left(\frac{r}{390r_g}\right)^{-1.5} .
\end{equation}
The transition radius is $r\approx 390r_g$.  This transition is
associated with the field line going from collimating to not
collimating (or slightly decollimating). This is along one mid-level
field line, and the coefficients and power laws are slightly different
for each field line.

The pitch angle shows that the field becomes toroidally dominated and
at large radius the magnetic loops have a pitch angle of $\lesssim
1^\circ$ by $10^3r_g$.  The small pitch-angle suggests that the flow
may be kink unstable.  The Kruskal-Shafranov criterion
($|B^{\hat{\phi}}/B^{\hat{r}}|\gtrsim
\sqrt{g_{\phi\phi}/g_{\theta\theta}}$), where
$\sqrt{g_{\theta\theta}}$ is approximately the vertical length of the
jet, is the standard kink instability criterion.  However, a
self-consistent treatment determined that a Poynting-dominated jet is
kink unstable only if both the Kruskal-Shafranov criterion is
satisfied {\it and} for a mostly radial flow that
$|B^{\hat{\phi}}/B^{\hat{r}}|> \sqrt{g_{\phi\phi}}/R_L$ \citep{tom01},
where $R_L=c/\Omega_F$ is the light cylinder radius and cylindrical
radius $R\approx \sqrt{g_{\phi\phi}}$ at large distances.  Their
result only strictly applies inside the light cylinder and for a
uniform field, but their result is suggestive of the general
result. The lower right panel of figure~\ref{jetstruct2} shows ${\rm
tan}^{-1} (R_L/\sqrt{g_{\phi\phi}})$ overlapping the pitch angle, and
so the flow is marginally stable to the kink instability. Thus, the
jet is {\it not} expected to be (violently) kink unstable if simulated
in 3D.

The above discussion implies that the radial magnetic field is well
fit by
\begin{equation}
B^{\hat{r}}\approx -\frac{c}{\Omega_F\sqrt{g_{\phi\phi}}} B^{\hat{\phi}} ,
\end{equation}
where we also find that $\Omega_F\approx \Omega_H/2$ is approximately
constant along a field line (also see, e.g., \citealt{mg04}), where
$\Omega_H=j/(2r_H)$ is the angular frequency of the black hole. The
marginal kink stability of our solution is also in concordance with
the fact that $\Omega_F\approx \Omega_H/2$ \citep{tom01}, since a
Poynting-dominated jet following such a rotation was shown to be
marginally kink stable.

As pinch instability-driven oscillations develop, this leads to
arbitrarily sized patches moving at arbitrary relative velocities as
required by any internal shock model.  A patch is simply defined as
a localized increase or decrease in various fluid quantities.  For
example, figure~\ref{jetstruct2} shows that at large radii that the
percent difference in the fluctuating part of the opening angle
depends on the hemisphere and varies between $10\%$ to $50\%$.
Comparing against each quantity smoothed over a logarithmic interval
of $d\log(r)=0.4$ for the hemisphere with less oscillations, the
typical percent difference of quantities along the length of the jet
along a mid-level field line is: $10\%$ for $\rho_0$, $b^2$,
$B^{\hat{r}}$, $B^{\hat{\phi}}$, $\Gamma$, and $u_\phi$ ; $20-50\%$
for $u$ ; and $30\%$ for $\Gamma_\infty$.  The other hemisphere with
larger $\theta_j$ oscillations has $\approx 5$ times the percent
difference compared to the first hemisphere.  For both hemispheres,
the largest several patches have an order unity percent difference.
The typical size of a patch is $\sim 100r_g$ to $\sim 10^3r_g$ in
the lab frame along the length of the jet.

The top two right panels of figure~\ref{jetstruct2} show the rest-mass
density per disk density and internal energy density per rest-mass
density.  The middle left panel shows the ratio of (twice) the
comoving magnetic energy density per unit rest-mass density. The
region within $r<6r_g$ is dominated by the floor model, and this is
where the mass is injected into the Poynting-jet.  This explains the
kink in the density profiles at $r\sim 6r_g$ and the plateau in
$b^2/\rho_0$.  For $r\gtrsim 10r_g$, the floor model is never
activated and the matter in the Poynting jet is self-consistently
evolved.  The inner-radial rest-mass density is well fit by
\begin{equation}\label{rhojet1}
\frac{\rho_0}{\rho_{0,disk}}\approx 1.5\times
10^{-9}\left(\frac{r}{120r_g}\right)^{-0.9}.~~~{\rm (inner)}
\end{equation}
The outer-radial rest-mass density is well fit by
\begin{equation}\label{rhojet2}
\frac{\rho_0}{\rho_{0,disk}}\approx 1.5\times
10^{-9}\left(\frac{r}{120r_g}\right)^{-2.2}.~~~{\rm (outer)}
\end{equation}
The transition radius is $r\approx 120r_g$.  Likewise, the
inner-radial internal energy density is moderately fit by
\begin{equation}\label{ujet1}
\frac{u}{\rho_{0,disk}c^2}\approx 4.5\times
10^{-9}\left(\frac{r}{120r_g}\right)^{-1.8}.~~~{\rm (inner)}
\end{equation}
The outer-radial internal energy density is moderately fit by
\begin{equation}\label{ujet2}
\frac{u}{\rho_{0,disk}c^2}\approx 4.5\times
10^{-9}\left(\frac{r}{120r_g}\right)^{-1.3}.~~~{\rm (outer)}
\end{equation}

For both the rest-mass and internal energy densities, the transition
radius is $r\approx 120r_g$, which is near the transition to a
superfast flow.  The location of this transition may be due to the
thickness of the disk, since the disk leads to collimation at small
radii.  A flow that collimates more leads to the fast surface closer
to the origin \citep{e93,begli94,tomimatsu1994}.  Hence, if a thick
disk is required for collimation, then a thick disk is required for a
superfast transition at small radii.

The internal energy density stabilizes in value once the
equipartition ``magnetic fireball'' has formed by $r\sim 10^3r_g$.
However, the simulation did not extend far enough to guarantee that
the equipartition extends for much larger radii.  The internal
energy may continue to grow, stay in equipartition, or decrease at
the expense of adiabatic expansion and acceleration of the jet.  The
resolution if this issue is left for future work.

\subsection{$\Theta$ Jet Structure}\label{thetajet}

\begin{figure*}
\begin{center}
\includegraphics[width=6.6in,clip]{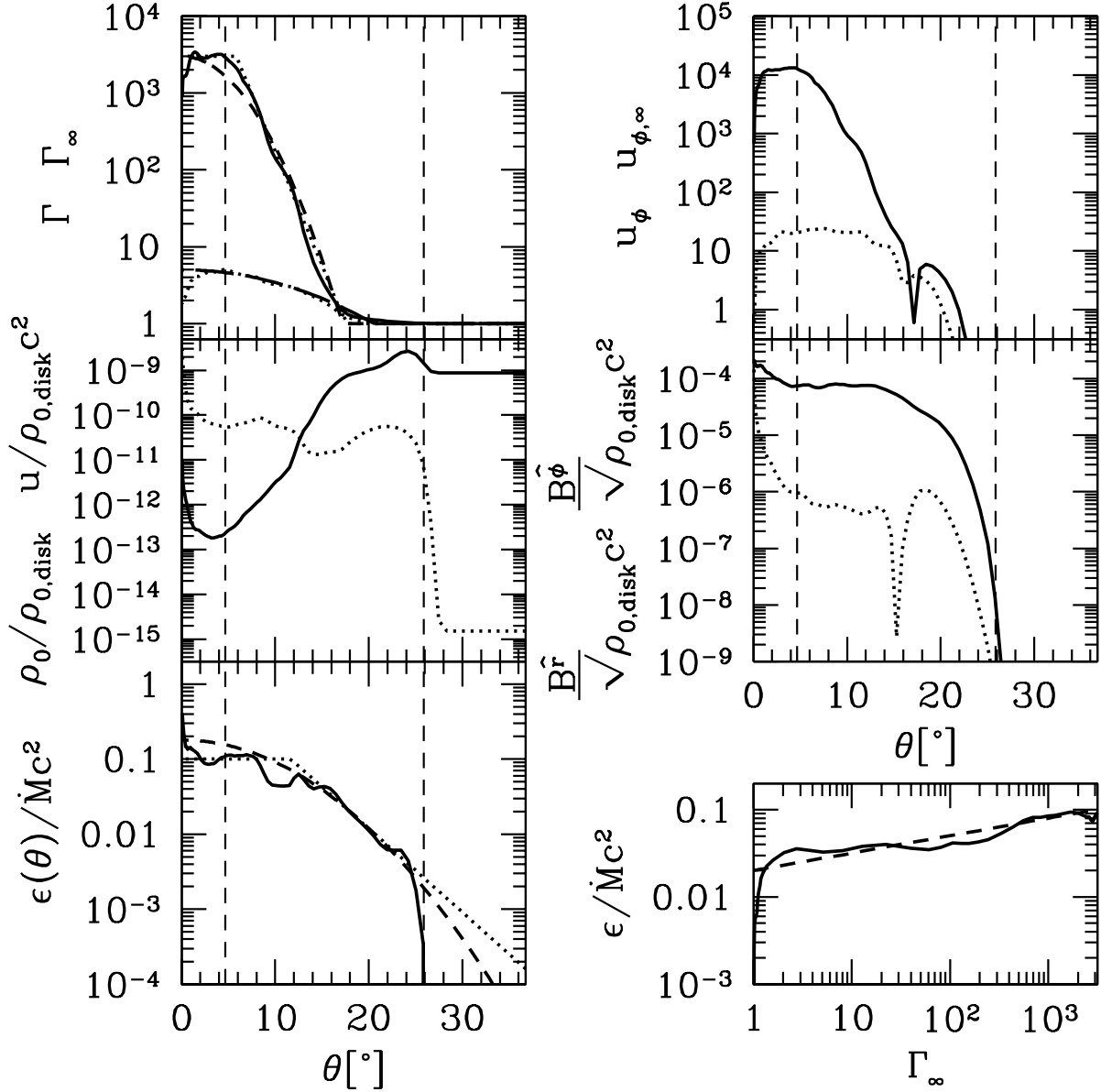}
\end{center}

\caption{ A $\theta$ cross-section of the jet at $r=5\times 10^3r_g$
showing the velocity, density, energy, and magnetic structure in
Gauss. $\Gamma_\infty$ and $\epsilon(\theta)/\dot{M}c^2$ have
overlapping Gaussian fits (dashed lines) and overlapping
uniform+exponential fits (dotted lines), where $\epsilon(\theta)\equiv
-r^2 T^r_t\approx r^2 \rho_0 u^r \Gamma_\infty$ (Boyer-Lindquist
coordinates).  A Gaussian fit to $\Gamma$ is shown as an overlapping
dotted line.  Lower right corner shows $\epsilon/\dot{M}c^2$ with
overlapping power-law fit (dashed line).  The magnetic field
components are in Boyer-Lindquist coordinates in an orthonormal
basis. Vertical dashed lines marks the outer jet and jet core.  Those
panels with two separate y-axes labels have the labels vertically
ordered by the function value near $\theta=0$. }
  \label{jetstruct}
\end{figure*}

Figure~\ref{jetstruct} shows the $\theta$ cross-section of the jet at
$r=5\times 10^3 r_g$ for the upper polar axis according to
figures~\ref{density}, ~\ref{densityzoom}, and~\ref{gammainf}.  This
is the location by which the jet has stabilized in time, where farther
regions are still dependent on the initial conditions.  The hot fast
``core'' of the jet includes $\theta\lesssim 5^\circ$ at this radius
and is marked by the left dashed vertical line.  Also, note that
$\theta\approx 8^\circ$ corresponds to the radius obtained for the
``mid-level'' field line shown for radial cross-sections in
figures~\ref{gammas},~\ref{jetstruct2}, and~\ref{thetaj}.

The right dashed vertical line marks the boundary between the last
field line that connects to the black hole.  This extended angular
distribution out to $\theta\lesssim 27^\circ$ is colder than the hot
core of the jet.  Beyond this region is the surrounding infalling
medium.  The disk wind has not reached such large radii.

The very inner core of the jet is denser and slower.  This feature is
simply a result of the fact that the Blandford-Znajek flux is $\propto
\sin^2{\theta}$.  Most of the Poynting energy flux is at the interface
between the jet and surrounding medium at small radii, where at larger
radii the jet internally evolves to collimate into a narrower inner
core where the energy flux is concentrated.

In most of our simulated models, the jet structure is nearly Gaussian.
The fiducial model has
\begin{equation}
\epsilon(\theta)= \epsilon_0 e^{-\theta^2/2\theta_\epsilon^2} ,
\end{equation}
where $\epsilon_0\approx 0.18\dot{M} c^2$ and $\theta_\epsilon\approx
8^\circ$.  The total luminosity per pole is $L_j\approx
0.023\dot{M}_0c^2$, where $10\%$ of that is in the ``core'' within a
half-opening angle of $5^\circ$ with the largest Lorentz factor within
the jet.  Also, $\Gamma_\infty$ is approximately Gaussian with
\begin{equation}
\Gamma_\infty(\theta)= \Gamma_{\infty,0} e^{-\theta^2/2\theta_\infty^2} ,
\end{equation}
where $\Gamma_{\infty,0}\approx 3\times 10^3$ and
$\theta_\infty\approx 4.3^\circ$.  The Lorentz factor is approximately
Gaussian with
\begin{equation}
\Gamma(\theta)= \Gamma_0 e^{-\theta^2/2\theta_\Gamma^2} ,
\end{equation}
where $\Gamma_0\approx 5$ and $\theta_\Gamma\approx 11^\circ$.

The jet structure can also be modelled by a small uniform core with
sharp exponential wings.  A good fit is
\begin{equation}
\epsilon(\theta)=0.1\dot{M} c^2~~~(\theta\le 11^\circ)
\end{equation}
and otherwise
\begin{equation}
\epsilon(\theta)=0.1\dot{M} c^2~e^{(-3(\theta/\theta_\epsilon-1))}
~~~(\theta\ge 11^\circ) ,
\end{equation}
where $\theta_\epsilon=11^\circ$. We find that
\begin{equation}
\Gamma_\infty=3\times 10^3~~~(\theta\le 5.8^\circ)
\end{equation}
and otherwise
\begin{equation}
\Gamma_\infty=3\times 10^3 e^{(-4(\theta/\theta_\infty-1))}~~~(\theta\ge 5.8^\circ) ,
\end{equation}
where $\theta_\infty=5.8^\circ$.

The bottom right corner more clearly demonstrates the fact that there
is a significant amount of energy flux at low Lorentz factors.  The
dependence roughly follows
\begin{equation}
\frac{\epsilon}{\dot{M}c^2}\approx 0.08
\left(\frac{\Gamma_\infty}{10^3}\right)^{1/5} .
\end{equation}

\subsection{Jet Characteristic Structure}\label{chars}

In this section the characteristic structure of the jet is considered.
First, the ideal MHD dispersion relation is given, e.g., in
\citet{gmt03} (there is a sign typo there), and summarized here.  In
the comoving frame, the dispersion relation is
\begin{eqnarray}
0  & = & D(k^\mu) \nonumber \\
& = & \omega \left(\omega^2 - k_a^2\right) \nonumber \\
& \times & \left(\omega^4 - \omega^2 \,
\left(
K^2 c^2_{ms} + c_s^2 k_a^2/c^2 \right) + K^2 c_s^2 k_a^2
\right),
\end{eqnarray}
where $k_a\equiv \kdv$, $c^2_{ms}\equiv (\va^2 + c_s^2 (1 -
\va^2/c^2))$ is the magnetosonic speed, $c_s^2 \equiv (\del (\rho +
u)/\del p)_s^{-1}$ is the relativistic sound speed, $\bva \equiv
\bB/\sqrt{\sE}$ is the relativistic \alf~velocity, $\sE \equiv b^2 +
w$, and $w \equiv \rho + u + p$. The invariant scalars defining the
comoving dispersion relation are $\omega = - k_\mu u^\mu$, $K^2 =
K_\mu K^\mu = k_\mu k^\mu+\omega^2$, where $K_\mu = P_{\mu\nu} k^\nu =
k_\mu - \omega u_\mu$ is the projected wave vector normal to the fluid
4-velocity, $\va^2 = b_\mu b^\mu/\sE$, and $\kdv = k_\mu
b^\mu/\sqrt{\sE}$.  The terms in the dispersion relation correspond
to, respectively from left to right, the zero frequency entropy mode,
the left and right going \alf~modes, and the left and right going fast
and slow modes.  The eighth mode is eliminated by the no-monopoles
constraint.  The dispersion relation gives the ingoing and outgoing
slow, \alf~and fast surfaces.

Energy can be extracted from the black hole if and only if the
\alf~point lies inside the ergosphere \citep{tak90}.  Optimal
acceleration of the flow by conversion of Poynting flux to kinetic
energy flux occurs beyond the outer fast surface
\citep{e93,begli94,tomimatsu1994}.  Other surfaces of interest
include: the horizon at $r_H \equiv 1+\sqrt{1-j^2}$ ; the ergosphere
at $r\equiv 1+\sqrt{1-(j\cos{\theta})^2}$ ; the coordinate basis light
surface in where $\sqrt{g_{\phi\phi}}=c/\Omega_F$, where
asymptotically $\sqrt{g_{\phi\phi}}=r\sin(\theta)$ is the Minkowski
cylindrical radius; and the surface in Boyer-Lindquist coordinates
where the toroidal field equals the poloidal field
($|B^{\hat{\phi}}|=|B^{\hat{r}}|$).  Finally, there is a stagnation
surface where the poloidal velocity $u^p=0$.  In a field confined jet
where no matter can cross field lines into the jet, and if the jet has
inflow near the black hole and outflow far from the black hole, then
this necessarily marks at least one location where rest-mass must be
created either by charge starving the magnetosphere till the
Goldreich-Julian charge density is reached \citep{gj69}, or pair
production rates sustains the rest-mass density to a larger value than
the Goldreich-Julian value.

\begin{figure}

\includegraphics[width=3.3in,clip]{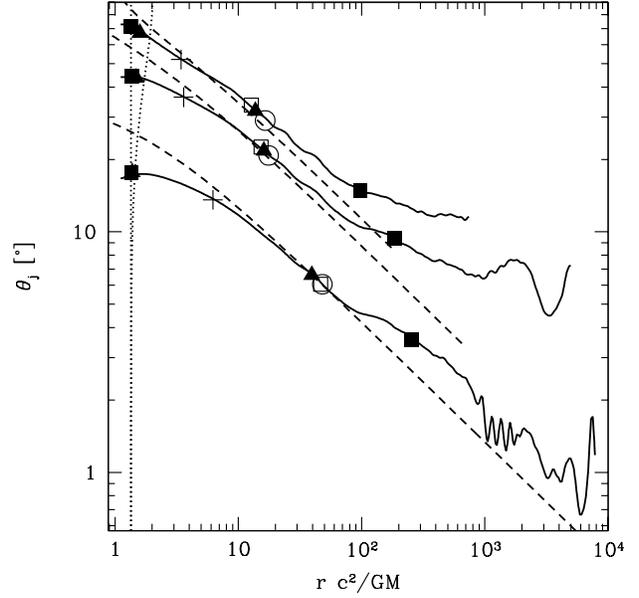}

\caption{ A radial cross-section of the Poynting-dominated jet along
three field lines showing the opening angle in degrees.  The field
lines are close to the polar axis, along a mid-level field line, and
close to the outer edge of the Poynting-dominated jet.  Overlapping
dashed lines are paraboloidal fits within the radial range that has a
similar collimation. Solid squares mark the ingoing/outgoing fast
points.  Solid triangles mark the ingoing/outgoing \alf~points.
Pluses mark the stagnation point.  Open squares mark where poloidal
and toroidal fields are equal.  Open circles mark the light
surface. Dotted lines mark the horizon and ergosphere. The field lines
at larger $\theta$ are shown for a shorter radius due to limitations
of a program used to extract a field line contour. }
  \label{thetaj}
\end{figure}

Figure~\ref{thetaj} shows $\theta_j$ as a function of radius for a
field line close to the polar axis, in the middle of the jet, and one
close to the outer edge of the jet.  Also shown is the
Blandford-Znajek paraboloidal solution given by equation 7.1 in
\citet{bz77}.  The three paraboloidal field lines correspond to
$X(r,\theta)/C={1.6,1.2,0.75}$ in that equation, where
$X(r,\theta)/C=2(1-\log(2))$ along the polar axis.  Note that at large
radii that a paraboloidal field has $\theta_j\propto r^{-1/2}$,
cylindrical has $\theta_j\propto r^{-1}$, and conical has
$\theta_j\propto r^{0}$. The ingoing fast point is nearly coincident
with the horizon. The stagnation points have time-dependent positions
that vary within approximately $\pm 30\%$ differences.

Notice from figure~\ref{thetaj} that the field near the horizon is
nearly monopolar.  After the stagnation point the field line becomes
nearly paraboloidal up to the outgoing \alf~point.  Beyond this point,
the field is unstable to pinch modes and some of the magnetic energy
is converted to thermal energy.  For the outer and mid-level field
lines, after the flow passes through the outgoing fast point, the flow
goes from nearly paraboloidal to nearly conical.  The inner-level
field line continues to follow a fairly paraboloidal collimation with
mild oscillations.  The magnetic surfaces are therefore paraboloidal
in the core of the jet and surrounded by conical surfaces, which is
quite similar to the assumed structure of the analytical model by
\citet{tt03}.

\begin{figure}
\includegraphics[width=3.3in,clip]{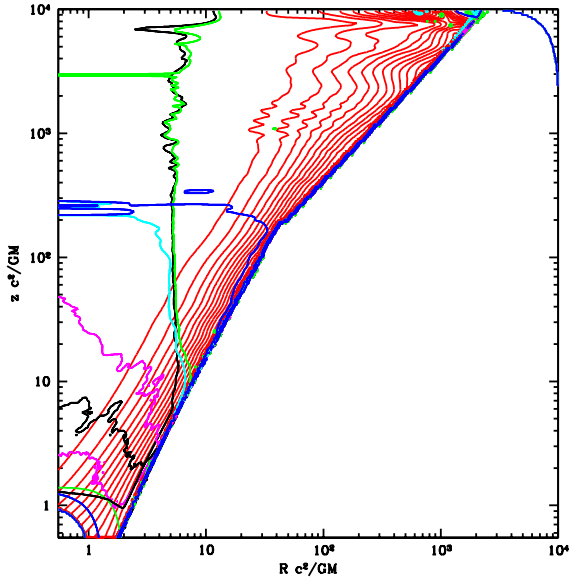}

\caption{Poynting-dominated jet characteristic (and other)
surfaces. Shows a log-log plot of one hemisphere of the time-averaged
flow at late time.  In such a log-log plot, $45^\circ$ lines
correspond to lines of constant $\theta$.  Lines of constant spherical
polar $r$ are horizontal near the $z$-axis and vertical near the
$R$-axis.  The field lines are shown as red lines.  From $r=0$
outwards: Blue\#1: horizon + ingoing-fast ; Cyan\#1: ingoing-\alf~;
Black\#1: $|B^{\hat{\phi}}|=|B^{\hat{r}}|$ ; Green\#1: ergosphere ;
Purple\#1: ingoing-slow ; Black\#2: stagnation surface where poloidal
velocity $u^p=0$ ; Purple\#2: outgoing-slow ; Cyan\#2: outgoing-\alf~;
Black\#3: $|B^{\hat{\phi}}|=|B^{\hat{r}}|$ again ; Green\#2: light
cylinder ; Blue\#2: outgoing-fast.  The disk and coronal regions have
been truncated with a power-law cutoff for $r\lesssim 100r_g$ and a
conical cutoff for larger radii.  Within $r\lesssim 10r_g$ the
plotting cutoff creates the appearance that the fast surface and other
lines terminate along the cutoff.}
  \label{wavespeeds}
\end{figure}

Figure~\ref{wavespeeds} shows the characteristic structure of the jet
for one polar axis.  The \alf~surface lies inside the ergosphere, as
required to extract energy \citep{tak90}.  Clearly the field lines
follow nearly a power-law until $r\sim 10^2 r_g$.  The stagnation
surface is time-dependent but stable.  Clearly the transition to a
supercritical (superfast) flow has occurred. After the fast surface,
the field lines stretch out and oscillate around a conical asymptote.
The fast surface near the polar axis is at $r\sim 250r_g$, where for
the other hemisphere it is at $r\sim 500r_g$. Other features are
quantitatively similar.

Despite the unsteady nature of the flow, the characteristic structure
is quite simple and relatively smooth in appearance.  This indicates
that the flow is mostly stationary.  The region near $r\sim 10^4 r_g$
is still determined by the initial conditions, which explains the
distorted appearance of the field lines and other artifacts related to
the flow being unsteady.

\subsection{Dependence on Surroundings}

The jet structure that emerges from the simulation is primarily due to
the internal evolution of the jet rather than by the interaction with
the surrounding medium. In particular, these GRMHD-based results are
insensitive to the two different models of the initial surrounding
medium.  One model is a surrounding infall of material and the other
is an ``evacuated'' exterior region, as discussed in sections~\ref{ic}
and~\ref{bc}.  The Poynting-dominated jet structure is negligibly
broader or narrower in the evacuated case due to magnetic confinement.
The disk wind itself also easily plows through the exterior region.

This is in stark contrast to the simulations of relativistic {\it
hydrodynamic} jets \citep{aloy00,zwm03,zwh04}, where they find that
the environment is primarily responsible for collimating the
hydrodynamic flow and hydrodynamic collimation shocks help diminish
the mixing between the jet and the surrounding medium.  In the GRMHD
case, we find that the confinement is mostly magnetic\footnote{There
is a difference between magnetic confinement and magnetic collimation.
Confinement refers to the magnetic field containing the mass of the
jet since there is no hydrodynamic mixing, while collimation refers to
modifications of the field line geometry along the jet.}.

The radial structure of the jet is also unaffected by the
environment. For example, one can compare the radial structure of
the density in the jet (equations~\ref{rhojet1} and~\ref{rhojet2})
to the density of the evacuated environment model of the
surroundings,
\begin{equation}
\frac{\rho_0}{\rho_{0,disk}}= 2\times
10^{-13}\left(\frac{r}{120r_g}\right)^{-2.7},~~~{\rm (Environment 1)}
\end{equation}
or to the pre-existing spherical infall environment model of the
surroundings,
\begin{equation}
\frac{\rho_0}{\rho_{0,disk}}= 8\times
10^{-8}\left(\frac{r}{120r_g}\right)^{-1.5} .~~~{\rm (Environment 2)}
\end{equation}
Even with the drastic difference in rest-mass and internal energy
densities, all the results of the Poynting-dominated jet are
quantitatively similar.  This is much different than the results
from hydrodynamic simulations mentioned above.

\section{Discussion}\label{discussion}

We find that the Poynting-dominated jet accelerates to $\Gamma\sim 10$
along nearly paraboloidal field lines.  The maximum terminal Lorentz
factor is $\Gamma_\infty\lesssim 10^3$, sufficiently relativistic to
account for any relativistic jet from GRBs, AGN, or x-ray binaries.
At large radii, about half of the energy in the jet is thermal.  Thus,
for optically thick flows that are radiation-dominated, thermal
adiabatic expansion is expected to lead to an extended acceleration
range far beyond $r\sim 10^4r_g$.  For optically thin flows that are
radiation-dominated, much of this energy can be directly released as
radiation.  In the limit that the radiative time-scale is much shorter
than the jet propagation time-scale, the terminal Lorentz factor is
limited to $\Gamma_\infty\sim \Gamma\sim 10$.

The part of the jet with the largest energy flux collimates from a
half-opening angle of $\theta_j\sim 60^\circ$ near the black hole to
$\theta_j\sim 5^\circ$ at large distances. The core of the jet follows
nearly paraboloidal field lines for all radii and accelerates with
approximately $\Gamma\propto r^{0.3}$ from $r\approx 10r_g$ out to
$r\approx 5\times 10^3 r_g$, after which the initial conditions
dominate the flow.  Thus the core moves with $\Gamma\approx 6$ and may
continue to accelerate since the field is still collimating by the
outer radius.  The middle and outer angular parts of the jet follow
nearly paraboloidal lines inside the fast surface and conical lines
outside the fast surface.  The middle and outer parts follow
$\Gamma\propto r^{0.45}$ from $r\approx 3r_g$ up to the point where
the flow becomes conical at $r\sim 10^2r_g - 10^3r_g$.  Thus these
parts move with $\Gamma\sim 10$.  The acceleration nearly follows the
result from analytic models with exactly paraboloidal field lines that
found $\Gamma(r)\propto r^{1/2}$ \citep{bn05}.  That the acceleration
found is slightly weaker than for paraboloidal lines is consistent
with the fact that the simulated field lines end up not as strongly
collimated as paraboloidal.

The jet structure is Gaussian-like, so there is a non-negligible
amount of energy at large angles out to $\theta_j\sim 27^\circ$.  This
may explain the observations of some GRBs that appear to have a narrow
ultrarelativistic component that produces the $\gamma$-rays and early
afterglow and a wide, mildly relativistic component that produces the
radio and optical afterglow \citep{berger03}. The provided fits to the
jet structures can be folded into various models, such as the
probability of observing polarised emission in Compton drag emission
models \citep{ghis00,lazzati2004} or whether Compton-drag models work.
Consider the region within $\theta\lesssim 27^\circ$.  This is an
expanded, relatively cold slow portion of the jet.  It is possible
that the gamma-ray burst photons are due to Compton drag from soft
photons emitted by this jet ``sheath'' dumping into the faster spine
\citep{bs87,ghis00,lazzati2004}.

\subsection{Toroidal Field Instabilities}

Internal shocks and magnetic instabilities have both been invoked to
explain emission from jets associated with GRBs, AGN, and x-ray binary
systems.  We find that the Poynting-dominated jet becomes unstable
beyond the \alf~surface and magnetic energy is converted to thermal
energy.  Unlike during reconnection, this process involves toroidal
field instabilities that lead to a conversion of magnetic to lateral
kinetic energy.  These oscillations drive waves into the jet that
steepen into shocks and lead to $\Gamma_\infty^{{\rm (MA)}}\sim
10^3$. This process is consistent with expectations of current-driven
instabilities \citep{e93,begelman1998,sikora2005}.  These
instabilities lead to slightly faster and slower moving patches of jet
material.  Thus, the variability within the jet is dominated by
toroidal field instabilities.

Toroidal field instabilities have been invoked to significantly lower
the BZ power output \citep{li2000}, although their estimate was based
upon the Kruskal-Shafranov criteria for kink instability, a
nonrelativistic model, and limited the BZ power output in the context
of an acausal astrophysical load at large distances.  At best their
estimate determines a limitation of the Poynting flux domination of
the jet, but not the total jet power or power emitted by the black
hole.  Kink instabilities have also been invoked to destroy the
coherent structure of jets (for a discussion, see, e.g.,
\citealt{appl96}), and such instabilities are often regarded as
sufficiently strong so that magnetic confinement is not possible
\citep{mj06}.  However, the Kruskal-Shafranov criteria for kink
instability is a necessary but not sufficient condition for kink
instabilities to occur, where rotation of the fluid or field lines can
stabilize the jet \citep{tom01}.

Blandford-Znajek type jet solutions naturally sit near marginal kink
stability \citep{tom01}.  Since we find that the self-consistently
simulated Poynting jet only marginally satisfies the necessary and
sufficient conditions for kink instability, the jet is not expected to
be (violently) kink unstable if simulated in full 3D.  This explains
why astrophysical jets can go to large distances even if satisfying
the Kruskal-Shafranov kink criteria.

Internal shocks are expected to generate some of the emission in GRBs,
AGN, and x-ray binaries.  In the simulated jet, the patches generated
by the toroidal field instabilities move at large relative Lorentz
factors, and so these patches can lead to internal shocks. However, by
the radius simulated, the Lorentz factor is not yet large enough for
the standard GRB internal shock model to be efficient
\citep{kob97,mes02,gcg03,piran2005}.  However, the energy provided by
the BZ-driven jet may be more than sufficient to allow for inefficient
shocks, since in section~\ref{thetajet} we found that $L\approx
0.023\dot{M}_0 c^2\sim 4\times 10^{51}\ergps$ for the collapsar model
rather than the canonical $\sim 10^{50}\ergps$ for cosmological GRBs.
The acceleration region and internal shock region was not simulated
since the dynamical range required is another $\sim 6$ orders of
magnitude in radius.

Toroidal field instabilities have been suggested to generate shocks
and high-energy emission in quasar jet systems \citep{sikora2005}. For
GRB systems, if magnetic dissipation continued beyond the $\gamma$-ray
photosphere at $r\sim 10^7 - 10^8 r_g$, then those shocks could be
directly responsible for the $\gamma$-ray emission \citep{lb01,lb03}.
This single magnetic dissipation model could unify GRB and Blazar
emission.  This paper does not resolve the emission mechanism, but the
results can be used to test the plausibility of emission models.

Some GRB models assume the Poynting energy to be converted into
radiation far from the collapsing star by internal dissipation (see,
e.g., \citealt{thompson94,mr97,sdd01,dren02,ds02,sbcp03,lpb03}) such
as driven by magnetic reconnection.  Kink instabilities have been
invoked to allow reconnection-driven magnetic energy to be converted
into thermal energy to accelerate the flow \citep{dren02,ds02},
although the reconnection rate assumed in those papers is not
supported by the GEM project on reconnection \citep{shay01}.  Some
simplified acceleration models assume that kink instabilities drive
reconnection that leads to acceleration \citep{gs06}.  However,
despite the simulated jet being marginally stable to current-driven
instabilities, significant internal dissipation occurs already by
$10^3r_g$.  Reconnection is not likely nor required. The shocks driven
by these moderate instabilities are an alternative to the reconnection
paradigm.

We find that after these toroidal field instabilities equalize the
magnetic and thermal energy at $r\sim 10^3r_g$ that the jet becomes
conical for the angular regions with $\theta_j\gtrsim 5^\circ$.
Consequently, this part of the jet ceases to magnetically accelerate
beyond this radius. In contrast, the core of the jet within $\theta_j
\approx 5^\circ$ remains paraboloidal along the entire jet out to
$r\sim 10^4r_g$. This core may continue to magnetically accelerate
beyond this radius. This suggests that toroidal field instabilities
play a crucial role in determining the opening angle and terminal
Lorentz factor of jets.

Theoretical studies of ideal MHD jets focused on cold jets and the
conversion of Poynting energy directly to kinetic energy (see, e.g.,
\citealt{lcb92,e93,begli94,tomimatsu1994,dd02,bn05}).  Indeed, much of
the work has focused on self-similar models, of which the only
self-consistent solution found is suggested to be R-self-similar
models \citep{vlah04,fo04}.  Unfortunately, such models result in only
cylindrical asymptotic solutions, which is apparently not what is
observed in AGN jets nor present in the simulations discussed in this
paper.  The previous section showed that some aspects of the jet are
nearly self-similar and that there is some classic ideal-MHD
acceleration occurring in this region.  However, cold ideal-MHD jet
models have no way of addressing the region where magnetic energy is
converted to thermal energy in shocks. This region also involves
relatively rapid variations in the flow, so stationary jet models
would have difficulty modelling this region.

\subsection{Simulations as Applied to AGN and X-ray Binaries}

Quantitatively all of these results can be rescaled by density in the
jet in order to approximately apply to any GRB model and to AGN and
x-ray binary systems.  This is because the jet has only reached about
$\Gamma\sim 10$, which is not too small or large a Lorentz factor, and
$\Gamma_\infty\propto 1/\rho_{0,jet}$.  Some astrophysical
implications of the jet simulations and results as applied to AGN are
described in the next several paragraphs, while a detailed discussion
of the results as applied to GRBs will be considered in a separate
paper.

The angular structure of the simulated jet may help explain many
aspects of observed jets.  The fast spine can lead to strong Compton
scattering of photons to higher energy, while the outer angular part
of the jet can lead to radio emission.  This process may be required
to explain high- and low- energy emission from TeV BL Lac objects and
radio galaxies \citep{ghis05} and may help explain high-energy
observations of blazars \citep{bl95,ghis96,chiab00}. The full opening
angle of $\sim 10^\circ$ of the simulated jet core also agrees with
the observations of the far-field jet in M87 \citep{junor99}.

The radial structure of the jet and the presence of shocks beyond the
\alf~surface suggest that synchrotron emission should be observed at
hundreds of gravitational radii from the black hole rather than near
the black hole.  This suggests the shock zone (or ``blazar zone'' for
blazars) should be quite extended between $r\sim 10^2r_g$ and up to
about $r\sim 10^4 r_g$ (see, e.g., \citealt{sikora2005}).  It is also
often assumed that if the jet is highly collimated that it is also
highly relativistic near the black hole, which would suggest
Comptonization of disk photons should produce clear spectral features
\citep{sm00}.  However, the jet may rather accelerate slowly but
collimate quickly, which is what we find.  There is an early transfer
of Poynting flux to kinetic energy flux leading up to about
$\Gamma\sim 5-10$ by about $r\sim 10^3 r_g$.  This is consistent with
the lack of observed Comptonization features in blazars (see, e.g.,
\citealt{sikora2005}).  At large distances the field becomes
toroidally dominated, but this does not necessarily contradict
observations of parallel fields in FRI sources
\citep{blandford00,sauty2002}.

Prior work suggested that the BZ power is insufficient to account for
Blazar emission \citep{mt03}, where they assumed that $\Gamma\sim 10$
and $\theta_j\sim 15^\circ$.  However, we find that the structure of
the Poynting-dominated jet is nontrivial.  The region with $\Gamma\sim
10$ is narrower with $\theta_j\sim 5^\circ$ and jet emission may be
dominated by shock accelerated electrons with thermal
$\Gamma^{\rm(MA)}\sim\Gamma\gtrsim 100$ with an extended high energy
tail.  This lowers the necessary energy budget of the jet enough to be
consistent with BZ power driving the jet.

The disk wind may play a crucial role independent of the jet.  For
example, relatively thin disks or slowly rotating black holes would
produce disk winds that could appear as ``aborted jets''
\citep{ghis04}.  The classical AGN unification models \citep{up95}
invoke a dominant role for the molecular torus and broad-line
emitting clouds, while the broad disk wind composed of magnetized
blob-like regions may significantly contribute to modifications and
in understanding the origin of the clouds
\citep{elvis2000,ernmfp04}.

Erroneous conclusions could be drawn regarding the jet composition
since there is actually a jet and disk wind, which could each have a
separate composition.  That there can be two separate relativistic jet
components makes it difficult to draw clear conclusions regarding the
composition \citep{guil83,cf93,lb96,sm00}.  Entrainment, which could
occur at large distances when the ideal MHD approximation breaks down,
also causes difficulties in isolating the ``proper'' jet component's
composition.  In cases where only electron-positron pair jets can be
supported, then this would indicate that disk winds are ruled out for
such sources, such as recently suggested for FRII sources
\citep{kt04}.

\subsection{Disk Winds vs. Poynting Jets in M87}

M87 is associated with the most widely studied AGN jet, so here we
discuss implications of our results for M87
specifically. \citet{junor99} and \citet{biretta99,biretta2002}
suggested that M87 slowly collimates from a full opening angle of
about $60^\circ$ near the black hole to $10^\circ$ at large distances.
However, some of their assumptions are too restrictive.  First, they
assumed the jet is always conical, which is apparent from figure 1 in
\citet{junor99}.  If the jet is not conical this can overestimate the
opening angle close to the core (i.e. perhaps $35^\circ$ is reasonable
all the way into the core).  Second, their beam size was relatively
large so that factors of $2$ error in the collimation angle are
possible. Third, they assumed the radio core is the location of the
black hole, while the inner-jet may be cold and only produce
synchrotron emission once shocks develop.  Finally, the assumption
that the jet is a single, slowly collimating jet remains the
prevailing view (see, e.g., \citealt{tb05,gracia05}).  GRMHD numerical
simulations of jet and disk wind formation suggest that a model with a
single jet is too simple of an interpretation of the observations.  If
there is a highly collimated relativistic Poynting jet surrounded by a
weakly collimated disk wind, then this would also fit their
observations without invoking slow collimation.

A form of the idea that winds are coincident with, and collimate, jets
has also been proposed by \citet{tb05} and applied to M87, but they
considered a model where the wind slowly collimates the jet in order
to fit observations.  Here we suggest that the observations may have
been misinterpreted due to the presence of two components: a
well-collimated relativistic cold Poynting jet and a mildly
relativistic disk wind.  We suggest the broader emission component
could be due to the disk wind.

More recent maps of the M87 jet formation region showed no ``jet
formation'' structure \citep{krich04}.  Thus, the structures seen
previously may be transient features, such as associated with
turbulent accretion disk producing a dynamic disk wind or associated
with time-dependent shocks produced in the jet far from the black
hole.

Measurements of the apparent jet speed in M87 reveal typically
$\Gamma\sim 1.8$ near the core while $\Gamma\sim 6$ at larger radii.
However, some core regions are associated with $\Gamma\sim 6$ that
rapidly fade \citep{biretta99}.  This is consistent with a
two-component outflow where the cold fast moving core of the jet is
only observed if it interacts with the surrounding medium, the slower
disk wind, or it undergoes internal shocks.

\subsection{Comparisons with Other Numerical Work}

Numerical models of jets continue to be studied using the hydrodynamic
approximation
\citep{mw99,aloy00,aloy02,scheck2002,zwm03,zwh04,aloy05}, despite the
presence of dynamically important magnetic fields in the accretion
disks that have been shown to lead to strongly magnetized jets
\citep{dhk03,mg04,dv05a}.  These hydrodynamic studies have suggested
that hydrodynamic mixing within the jet may lead to observational
signatures in the emission variability.  Other models suggested that
such hydrodynamic mixing leads to an inability to clearly distinguish
the jet composition as either leptonic or baryonic
\citep{scheck2002}. We suggest magnetic instabilities due to pinch or
kink modes dominate the variability rather than hydrodynamic
instabilities, which are known to be quenched by magnetic confinement
(see, e.g., \citealt{clarke86,lind89,ac92,rosen99}).  The magnetic
confinement severely limits the mixing between the jet and the
environment.

Hydrodynamic studies have suggested that shocks driven into the
surrounding environment collimate the jet \citep{aloy00,zwm03,zwh04}.
We find that the jet structure is negligibly broader for models with a
negligible surrounding environment.  This is because the jet is
magnetically confined and collimated at large radii. As applied to
GRBs, our results suggest that the presence of an extended stellar
envelope plays no role in the jet dynamics except to provide the disk
with material.  However, the jet plays a significant role in modifying
the stellar envelope by sending lateral shocks through the star.

Similar studies have also suggested that variations in the Lorentz
factor are due to hydrodynamic effects \citep{aloy02,aloy05}, but we
find that variations in the Lorentz factor are due to magnetic
instabilities.  Unlike their models, our models generate a relatively
simple jet structure that can be fit by a Gaussian or top-hat +
exponential wings. Once the shock heating generates an equilibrium
magnetic fireball, the internal thermal support and toroidal magnetic
confinement keep the jet stable and conical out to large distances.
For an optically thick flow like in GRB jets, the flow should convert
the thermal energy to kinetic energy in adiabatic expansion and lead
to $\Gamma\sim 10^3$.

For hydrodynamic models of GRBs, energy is injected to model the
annihilation of neutrinos \citep{aloy00,zwm03,zwh04}.  They assumed a
highly relativistic jet is formed early near the black hole and they
injected the matter with a large enthalpy per baryon of about
$u/\rho_0c^2\sim 150$.  They tuned the injected energy to have their
results agree with observations \citep{frail01}. Other models with a
disk and jet find that $u/\rho_0c^2\sim 10$, which is suggested to
imply that $\Gamma_\infty\sim 10$, insufficient for GRBs \citep{mw99},
see their figure 28.  In other work we suggested that the injected
energy per baryon is only $u/\rho_0c^2\lesssim 20$ unless
super-efficient neutrino mechanisms are invoked
\citep{mckinney2005b,rs2005}.  We also found that the energy released
is larger than observed, suggesting an fairly inefficient generation
of $\gamma$-rays.  We find an energetically dominant, lower $\Gamma$
jet component that may explain x-ray flashes.  In our case the
baryon-contamination problem is avoided by magnetic confinement of the
jet against baryons from the disk.  We find $\Gamma_\infty\sim
100-1000$, which is much larger than they found and this may avoid the
compactness problem.  The difference between their and our model is
the presence of a magnetic field and a rotating black hole, which
together drive the BZ-effect and a stronger evacuation of the polar
jet region.

MHD simulations of jet propagation have been performed that correspond
to weakly magnetized AGN jets, where $b^2/(2u)\lesssim 3.3$ and
$b^2/\rho_0<0.2$ \citep{leismann2005}.  Our study suggests that this
region may only be important for AGN Poynting-dominated jets at scales
$r\gg 10^4r_g$ once the jet becomes matter-dominated or may apply to
the disk outflow.

Nonrelativistic MHD \citep{proga03,kato04} and prior GRMHD simulations
\citep{mizuno04b,kom05} of jets showed a slow jet with $v\sim
0.2-0.3c$, insufficient to explain most jets from GRBs and AGN.
\citet{proga03} suggested that their Poynting-dominated jet has
$\Gamma_\infty\lesssim 10$, but they were unable to follow the
relativistic jet using the nonrelativistic equations of motion.  The
previous GRMHD simulations created slow jets because of the ``floor
model'' in the polar regions, the short time of integration, and the
limited outer radius of the computational box.

\citet{dv05b} used a nonconservative numerical method to evolve a
fully relativistic, black hole mass-invariant model, and showed a jet
with hot blobs moving with $\Gamma\sim 50$. A primary result in
agreement with the results here is that a patchy or pulsed ``magnetic
fireball'' is produced.  This suggests that the development of a
``magnetic fireball'' is not an artifact of the numerical
implementation but is a result of shock heating.  We also agree in
finding that the core of the jet is hot and fast and is surrounded by
a cold slow flow.  One difference is that they say they seem to have
found that the flow is cylindrically collimated by $r\gtrsim 300r_g$,
while we find nearly power-law collimation until $r\sim 10^3 r_g$ and
an oscillatory conical asymptote beyond. They suggested that temporal
variability is due to injection events near the black hole, while we
suggest it is due to pinch (or perhaps kink) instabilities at
$r\gtrsim 10^2r_g$.

They found a larger value of $\Gamma$ at smaller radius than we find.
Indeed, one should wonder why their black hole mass-invariant model
applies to only GRBs and not AGN or x-ray binaries.  This is a result
of their {\it much} lower ``floor'' density of $\rho_0\sim
10^{-12}\rho_{0,disk}$, which is far too low to be consistent with the
self-consistent pair-loading \citep{pwf99} or neutron diffusion
loading \citep{le2003} of the jet. Also, because they used a constant
density floor, the jet region at large radius must be additionally
loaded with an arbitrary amount of rest-mass.  The typical ``floor''
model adds rest-mass into the comoving frame, but this artificially
loads the jet with extra mass moving at high $\Gamma$.  As applied to
GRBs, we suggest that rather than the acceleration occurring within
$r\lesssim 700r_g$ of their simulation, that acceleration occurs much
farther away before the emitting region at $r\lesssim 10^9 r_g$.  In
their $\Gamma\sim 50$ knots the gas has $10^3\lesssim
u/\rho_0c^2\lesssim 10^6$.  This implies that their actual terminal
Lorentz factor is on the order of $10^5\lesssim \Gamma_\infty\lesssim
10^7$, which would imply that the external shocks occur before
internal shocks could occur.

\subsection{Limitations}

As has been pointed out by \citet{kom05}, {\it nonconservative} GRMHD
schemes often overestimate or underestimate the amount of thermal
energy produced.  All numerical models suffer from some numerical
error.  Shock-conversion of magnetic energy to thermal energy is
modelled by the {\it conservative} scheme in HARM in the perfect
magnetic fluid approximation with total energy conserved exactly.
However, our numerical model may still have overestimated the amount
of magnetic dissipation in shocks, and a more accurate calculation may
more slowly convert magnetic energy to thermal energy.  However, we
expect that toroidal field instabilities drive efficient magnetic
dissipation as shown in the numerical results presented here.  It is
also encouraging that similar result are found in 3D nonrelativistic
simulations \citep{ocp03} and maybe in the 2D relativistic simulations
of \citet{dv05b} (they did not say why their jet generates hot blobs).

Confidence in the numerical results is obtained by code testing
\citep{gmt03}, comparisons with related analytical models
\citep{mg04}, and convergence testing the specific model being
studied.  For the fiducial model simulated in this paper, we used both
higher and lower order spatial and temporal reconstructions and used a
resolution smaller by a factor of two.  The results described in the
paper are independent of these numerical changes.

Unlike GRB neutrino-dominated disks, AGN and X-ray binaries may have
accretion disks with a wide variety of $H/R$ near the black hole,
which could lead to a wide to narrow opening angle of the jet and disk
wind.  This broadness of the jet and wind may affect the classical
viewing-angle-dependent unification models \citep{sauty2002}.  Only a
self-consistent radiative GRMHD calculation can determine the disk
thickness.  The quantitative conclusions in this paper regarding the
collimation angle assumed $H/R\sim 0.2$ near the black hole, while
$H/R\sim 0.9$ (ADAF-like, \citealt{ny95}) is perhaps more appropriate
for such systems.  The sensitivity of these results to $H/R$ is left
for future work.

The simulations used a field geometry that was initially moderately
organized, although there was no net field.  As discussed in the
introduction and in the floor model section, moderate changes in the
initial geometry do not affect the results \citep{mg04}.  However,
accretion of a highly irregular (tangled) field would be quite
different.  An organized field may not develop around the black hole,
and so disk material would mass-load the jet.  In this case, the jet
just becomes an extension of the disk wind \citep{mg04}.  However, the
existence of a mostly uniform field threading the disk arises
naturally during core-collapse supernovae and NS-BH collision debris
disks.  In AGN and stellar wind- capture x-ray binary systems, the
accreted field may be uniform over long time scales
\citep{narayan2003,pu05}.  Roche-lobe overflow x-ray binaries,
however, may accrete quite irregular field.  The field geometry that
arrives at the black hole, after travelling from the source of
material (molecular torus, star(s), etc.) to the black hole horizon,
should depend sensitively on the reconnection physics.

In order to evolve for a longer time than simulated in this paper,
other physics must be included.  For long-term evolution of a GRB
model, one must include disk neutrino cooling, photodisintegration of
nuclei, and a realistic equation of state.  If one wishes to track
nuclear species evolution, a nuclear burning reactions network is
required.  For the neutrino optically thick region of the disk,
radiative transport should be included.  The self-gravity of the star
should be included to evolve the core-collapse.  This includes a
numerical relativity study of the collapse of a rotating magnetized
massive star into a black hole (see review by \citealt{st03}, \S 4.3).
The jet should be followed through the entire star and beyond
penetration of the stellar surface \citep{aloy00,zwm03,zwh04}.

As applied to all black hole accretion systems, some other limitations
of the numerical models presented include the assumption of
axisymmetry, ideal MHD, and a nonradiative gas.

The assumption of axisymmetry is not crucial for the basic structure
of the inner jet region since our earlier results \citep{mg04} are in
quantitative agreement with 3D results \citep{dhk03}.  The primary
observed limitation of axisymmetry appears to be the decay of
turbulence \citep{cowling34}, which we attempted to avoid by requiring
a resolution that gives quasi-steady turbulence for much of the
simulation.  Also, the jet at large distances has already formed by
the time turbulence decays, and by that time the jet at large radius
is not in causal contact with the disk.

When the toroidal field dominates the poloidal field, eventually $m=1$
kink instabilities and higher modes may appear in 3D models (see,
e.g., \citealt{nakamura2001,ocp03,nakamura2004}).  Thus our 2D
axisymmetric models may have underestimated the amount of oscillation
in the flow and the conversion of Poynting flux to enthalpy flux.  On
the contrary, rotating Poynting-dominated jets may be marginally
stable to kink instabilities \citep{tom01}, as suggested by our
results.

The decay of turbulence due to the limitation of axisymmetry may also
{\rm limit} the efficiency of the Blandford-Znajek process.  The
magnetic arrested disk (MAD) model suggests that any accretion flow
that accumulates a large amount of magnetic flux near the black hole
eventually halts the accretion flow and builds a black hole
magnetosphere \citep{igumenshchev2003,narayan2003}.  This implies that
the efficiency of extracting energy could be higher.

We have neglected high-energy particle and radiative processes in the
numerical simulations.  The floor model of the minimum allowed
rest-mass density and internal energy density is ad hoc and chosen so
that a fast jet emerges.  Future work should include a model of pair
creation and pair annihilation in order to simulate self-consistent
mass-loading of the Poynting-dominated jet.  A self-consistent model
of the mass-loading can at least determine the density that should be
present near the black hole around the poles.  However, accurate
determination of the Lorentz factor requires at least radiative
transfer and Comptonization to model the radiative reaction within the
jet that can slow or speed the jet.

For AGN and x-ray binaries, the radiatively inefficient disk
approximation, which assumes electrons couple weakly to ions, may not
hold.  If the electrons and ions eventually couple near the black
hole, then the disk might collapse into an unstable magnetically
dominated accretion disk (MDAF) \citep{meier2005}.  Like the MAD
model, this might drastically alter the results here, although it is
uncertain whether jets are actually produced under the conditions
specified by the MDAF model.

The single-fluid, ideal MHD approximation breaks down under various
conditions, such as during the quiescent output of AGN and x-ray black
hole binaries, where a two-temperature plasma may form near the black
hole as ions and electrons decouple (see, e.g., \citealt{ny95}).  In
some AGN and x-ray binary systems, the gas may no longer act like a
fluid and nonthermal Fermi acceleration can produce a jet in a shear
acceleration region between the disk and corona \citep{sub99}.
Resistivity plays a role in current sheets where reconnection events
may generate flares as on the sun, such as possibly observed in Sgr~A
\citep{genzel2003}.  Finally, radiative effects may introduce
dynamically important instabilities in the accretion disk (e.g.
\citealt{gam98,bs03}).

\subsection{Future Work}

As applied to the collapsar model and other GRB models, future
calculations will include a realistic equation of state;
photodisintegration of nuclei; general relativistic ray-tracing
neutrino transport (similar to, e.g., \citealt{bl05}); neutrino
cooling (similar to, e.g., \citealt{kohri2005}); general relativistic
accounting for the neutrino annihilation; neutrino Comptonization;
simplified photon transport ; and photon Comptonization.  The
numerical calculations will also be performed in 3D.

As applied to AGN and x-ray binaries, future calculations will
include general relativistic photon transport (similar to, e.g.,
\citealt{bl05}); photon cooling; general relativistic accounting for
the photon annihilation in the Poynting-dominated jet; photon
Comptonization; and thermal and nonthermal synchrotron emission. The
numerical calculations will also be performed in 3D.

The same type of calculation can be performed for systems harboring
neutron stars, and this is a natural extension of recent force-free
work \citep{mckinney2006a,mckinney2006b}.


\section{Summary}\label{conclusions}

A GRMHD code, HARM, was used to evolve black hole accretion disk
models.  Our work extends the results of previous GRMHD numerical
models by studying the Poynting-dominated jet in more detail and
studying the jet out to $r\sim 10^4 r_g$ till $t\sim 10^4t_g$.

Basic results and conclusions include:
\begin{enumerate}
  \item Poynting jet reaches $\Gamma\sim 10$ following nearly
  paraboloidal field lines and nearly $\Gamma\propto r^{1/2}$.  Beyond
  $r\sim 10^3r_g$, the jet has a conical outer angular part and a
  nearly paraboloidal core.
  \item Peak energy flux within the Poynting jet goes from $\theta_j\sim
  60^\circ$ near the black hole to $\theta_j\sim 5^\circ$ at large
  distances.
  \item Energy structure of the Poynting jet is Gaussian with a
  half-width of $\theta_0\approx 8^\circ$.
  \item Poynting jet has a core with flat energy flux within $\theta_j\approx 5^\circ$.
  \item Extended slower Poynting jet component with $\theta_j\approx
  27^\circ$.
  \item Poynting jet variability is dominated by toroidal field instabilities.
  \item Poynting jet is marginally kink stable.
  \item Poynting flux is shock-converted into enthalpy flux beyond
  \alf~surface in an extended ``shock zone.''
  \item Maximum terminal Lorentz factor is $\Gamma_\infty\lesssim 10^3$.
  \item Poynting flux and enthalpy flux come into equipartition by
  $r\sim 10^3r_g$.
  \item For radiation-dominated conditions, optically thick flows can
  tap thermal energy in adiabatic acceleration, while optically thin
  flows can lose thermal energy as radiation.
  \item Disk wind has collimated edge (near the Poynting jet) with
  $\Gamma\lesssim 1.5$.
  \item Full disk wind is broad even near the black hole, and this may
  account for significant spatially-broad emission from near AGN
  cores, such as in M87.
\end{enumerate}

\section*{Acknowledgments}

This research was supported by NASA-Astrophysics Theory Program grant
NAG5-10780 and a Harvard CfA Institute for Theory and Computation
fellowship. I thank Avery Broderick for an uncountable number of
inspiring conversations.  I also thank Charles Gammie, Brian Punsly,
Amir Levinson, and Ramesh Narayan, with whom each I have had inspiring
conversations.  I thank Amir Levison, Jonathan Granot, Maxim Lyutikov,
Brian Punsly, Rob Fender, and Beskin Vasily for comments and
discussions on the draft version of this paper.  I thank Scott Noble
for providing his highly efficient and accurate primitive variable
solver.  I thank Xiaoyue Guan for providing her implementation of
parabolic interpolation.






\label{lastpage}


\begin{thebibliography}{99}





%
%
%
\bibitem[Abramowicz, Jaroszinski, \& Sikora(1978)]{ajs78} Abramowicz,
        M., Jaroszinski, M., \& Sikora, M. 1978, \aap, 63, 221
\bibitem[Akiyama, Wheeler, Meier, \& Lichtenstadt(2003)]{awml03}
Akiyama, S., Wheeler, J.~C., Meier, D.~L., \& Lichtenstadt, I.\
2003, \apj, 584, 954 
\bibitem[Aloy et al.(2000)]{aloy00} Aloy, M. A., Muller, E., Ibanez, J. M., Marti, J. M., \& MacFadyen, A. I.  2000, \apjl, 531,  L119

\bibitem[Aloy et al.(2002)]{aloy02} Aloy, M.-A., Ib{\' a}{\~
n}ez, J.-M., Miralles, J.-A., \& Urpin, V.\ 2002, \aap, 396, 693


\bibitem[Aloy et al.(2005)]{aloy05} Aloy, M.~A., Janka, H.-T.,
Muller, E.\ 2005, \aap, 436, 273


\bibitem[Anile(1989)]{anile} Anile, A.M. 1989,  Relativistic Fluids and
        Magneto-fluids, (New York: Cambridge Univ. Press)

\bibitem[Appl \& Camenzind(1992)]{ac92} Appl, S., \&
Camenzind, M.\ 1992, \aap, 256, 354


\bibitem[Appl(1996)]{appl96} Appl, S.\ 1996, \aap, 314, 995



\bibitem[Balbus \& Hawley(1991)]{bh91} Balbus, S.~A.~\& Hawley, J.~F.\
  1991, \apj, 376, 214
\bibitem[Begelman et al.(1984)]{begelman84} Begelman, M.~C.,
Blandford, R.~D., \& Rees, M.~J.\ 1984, Reviews of Modern Physics,
56, 255 
\bibitem[Begelman \& Sikora(1987)]{bs87} Begelman, M.~C., \&
Sikora, M.\ 1987, \apj, 322, 650 
\bibitem[Begelman \& Li(1994)]{begli94} Begelman, M.~C., \& Li,
Z.\ 1994, \apj, 426, 269 
\bibitem[Begelman et al.(1994)]{brs94} Begelman, M.~C., Rees,
M.~J., \& Sikora, M.\ 1994, \apjl, 429, L57 
%
\bibitem[Begelman(1998)]{begelman1998} Begelman, M.~C.\ 1998, \apj,
493, 291 

\bibitem[Berger et al.(2003)]{berger03} Berger, E., et al.\
2003, \nat, 426, 154


\bibitem[Beskin(1997)]{beskin97} Beskin, V.~S.\ 1997, Uspekhi
Fizicheskikh Nauk, 40, 659


\bibitem[Beskin \& Nokhrina(2005)]{bn05} Beskin, V.~S., \&
Nokhrina, E.~E.\ 2005, ArXiv Astrophysics e-prints,
arXiv:astro-ph/0506333
%
%

\bibitem[Biretta et al.(1999)]{biretta99} Biretta, J.~A., Sparks,
W.~B., \& Macchetto, F.\ 1999, \apj, 520, 621 
\bibitem[Biretta et al.(2002)]{biretta2002} Biretta, J.~A., Junor,
W., \& Livio, M.\ 2002, New Astronomy Review, 46, 239 
\bibitem[Bisnovatyi-Kogan \& Ruzmaikin(1976)]{br76}
Bisnovatyi-Kogan, G.~S., \& Ruzmaikin, A.~A.\ 1976, \apss, 42, 401
\bibitem[Blaes \& Socrates(2003)]{bs03} Blaes, O., \&
Socrates, A.\ 2003, \apj, 596, 509 
\bibitem[Blandford(1976)]{blandford76} Blandford, R.~D.\ 1976,
\mnras, 176, 465 
%
%
\bibitem[Blandford \& Znajek(1977)]{bz77} Blandford, R.~D.~\& Znajek,
  R.~L.\ 1977, \mnras, 179, 433

\bibitem[Blandford(2000)]{blandford00} Blandford, R.~D.\ 2000,
Astronomy, physics and chemistry of $H^{+}_{3}$, 358, 811



\bibitem[Blandford \& Levinson(1995)]{bl95} Blandford,
R.~D., \& Levinson, A.\ 1995, \apj, 441, 79 
\bibitem[Bogovalov \& Tsinganos(2005)]{bt05} Bogovalov, S.,
\& Tsinganos, K.\ 2005, \mnras, 357, 918 

\bibitem[Broderick \& Loeb(2005)]{bl05} Broderick, A.~E., \&
Loeb, A.\ 2005, \mnras, 363, 353



\bibitem[Celotti \& Fabian(1993)]{cf93} Celotti, A., \&
Fabian, A.~C.\ 1993, \mnras, 264, 228 
\bibitem[Chiaberge et al.(2000)]{chiab00} Chiaberge, M.,
Celotti, A., Capetti, A., \& Ghisellini, G.\ 2000, \aap, 358, 104 

\bibitem[Clarke et al.(1986)]{clarke86} Clarke, D.~A., Norman,
M.~L., \& Burns, J.~O.\ 1986, \apjl, 311, L63


\bibitem[Coburn \& Boggs(2003)]{cb03} Coburn, W.~\& Boggs,
S.~E.\ 2003, \nat, 423, 415 
\bibitem[Colella \& Woodward(1984)]{colella84} Colella, P, \& Woodward, P., 1984, JCP, 54, 174
\bibitem[Cowling(1934)]{cowling34} Cowling, T.~G.\ 1934, \mnras, 94, 768
\bibitem[Cui et al.(1998)]{cui98} Cui, W., Zhang, S.~N., \&
Chen, W.\ 1998, \apjl, 492, L53

\bibitem[Daigne \& Drenkhahn(2002)]{dd02} Daigne, F., \&
Drenkhahn, G.\ 2002, \aap, 381, 1066 
\bibitem[De Villiers, Hawley, \& Krolik(2003)]{dhk03} De
Villiers, J., Hawley, J.~F., \& Krolik, J.~H.\ 2003, \apj, 599, 1238
\bibitem[De Villiers et al.(2005a)]{dv05a} De Villiers, J.,
Hawley, J.~F., Krolik, J.~H., \& Hirose, S.\ 2005, \apj, 620, 878 
\bibitem[De Villiers et al.(2005b)]{dv05b} De Villiers, J.,
Staff, J., \& Ouyed, R.\ 2005, astro-ph/0502225 
\bibitem[Di Matteo et al.(2005)]{dimat05} Di Matteo, T.,
Springel, V., \& Hernquist, L.\ 2005, \nat, 433, 604 
\bibitem[Drenkhahn(2002)]{dren02} Drenkhahn, G. 2002, \aap, 387, 714
\bibitem[Drenkhahn \& Spruit(2002)]{ds02} Drenkhahn, G. \& Spruit, H. C.  2002,  \aap, 391, 1141
\bibitem[Duncan \& Thompson(1992)]{dt92} Duncan, R.~C.~\&
Thompson, C.\ 1992, \apjl, 392, L9 
\bibitem[Eichler(1993)]{e93} Eichler, D.\ 1993, \apj, 419,  111
\bibitem[Elvis(2000)]{elvis2000} Elvis, M.\ 2000, \apj, 545, 63
\bibitem[Elvis et al.(2004)]{ernmfp04} Elvis, M., Risaliti, G.,
Nicastro, F., Miller, J.~M., Fiore, F., \& Puccetti, S.\ 2004,
\apjl, 615, L25 
\bibitem[Fabian et al.(2002)]{fabianvaughan2002} Fabian, A.~C., et al.\
2002, \mnras, 335, L1 
\bibitem[Fender(2003a)]{fender2003a} Fender, R.\ 2003a, ArXiv
Astrophysics e-prints, arXiv:astro-ph/0303339 
\bibitem[Fender(2003b)]{fender2003b} Fender, R.~P.\ 2003b, \mnras,
340, 1353


\bibitem[Fender \& Belloni(2004)]{fb04} Fender, R., \&
Belloni, T.\ 2004, \araa, 42, 317 

\bibitem[Fendt \& Ouyed(2004)]{fo04} Fendt, C., \& Ouyed,
R.\ 2004, \apj, 608, 378


\bibitem[Fishbone \& Moncrief(1976)]{fm76} Fishbone, L.G., \& Moncrief,
        V. 1976, \apj, 207, 962
\bibitem[Frail et al.(2001)]{frail01} Frail, D.~A., et al.\
2001, \apjl, 562, L55 
\bibitem[Gammie(1998)]{gam98} Gammie, C.~F.\ 1998, \mnras,
297, 929 
\bibitem[Gammie~et~al.(2003a)]{gmt03} Gammie, C. F., McKinney, J. C.,
\& G$\acute{a}$bor T$\acute{o}$th 2003, \apj, 589, 444 
\bibitem[Gammie, Shapiro, \& McKinney(2004)]{gsm04} Gammie,
C.~F., Shapiro, S.~L., \& McKinney, J.~C.\ 2004, \apj, 602, 312 
\bibitem[Gammie(2004)]{gam04} Gammie, C.~F.\ 2004, \apj, 614, 309
\bibitem[Genzel et al.(2003)]{genzel2003} Genzel, R., Sch{\"
o}del, R., Ott, T., Eckart, A., Alexander, T., Lacombe, F., Rouan,
D., \& Aschenbach, B.\ 2003, \nat, 425, 934 
\bibitem[Gierli{\' n}ski \& Done(2004)]{gd04} Gierli{\'
n}ski, M., \& Done, C.\ 2004, \mnras, 347, 885 
\bibitem[Ghirlanda et al.(2003)]{gcg03} Ghirlanda, G.,
Celotti, A., \& Ghisellini, G.\ 2003, \aap, 406, 879 
%
\bibitem[Ghisellini et al.(1993)]{ghis93} Ghisellini, G.,
Padovani, P., Celotti, A., \& Maraschi, L.\ 1993, \apj, 407, 65 
\bibitem[Ghisellini \& Madau(1996)]{ghis96} Ghisellini, G., \&
Madau, P.\ 1996, \mnras, 280, 67 
\bibitem[Ghisellini et al.(2000)]{ghis00} Ghisellini, G.,
Lazzati, D., Celotti, A., \& Rees, M.~J.\ 2000, \mnras, 316, L45 
\bibitem[Ghisellini \& Celotti(2001)]{gc01} Ghisellini, G.,
\& Celotti, A.\ 2001, \mnras, 327, 739 
\bibitem[Ghisellini et al.(2004)]{ghis04} Ghisellini, G.,
Haardt, F., \& Matt, G.\ 2004, \aap, 413, 535 
\bibitem[Ghisellini et al.(2005)]{ghis05} Ghisellini, G.,
Tavecchio, F., \& Chiaberge, M.\ 2005, \aap, 432, 401 
\bibitem[Giannios \& Spruit(2006)]{gs06} Giannios, D., \&
Spruit, H.~C.\ 2006, ArXiv Astrophysics e-prints,
arXiv:astro-ph/0601172


\bibitem[Ghosh \& Abramowicz(1997)]{ga97} Ghosh, P.~\&
Abramowicz, M.~A.\ 1997, \mnras, 292, 887 
\bibitem[Goldreich \& Julian(1969)]{gj69} Goldreich, P., \&
Julian, W.~H.\ 1969, \apj, 157, 869 
\bibitem[Goodman(1997)]{goodman1997} Goodman, J.\ 1997, New
Astronomy, 2, 449 

\bibitem[Gracia et al.(2005)]{gracia05} Gracia, J., Tsinganos,
K., \& Bogovalov, S.~V.\ 2005, \aap, 442, L7




\bibitem[Granot \& K{\"o}nigl(2003)]{gk03} Granot, J. \& K{\"o}nigl, A.\
2003, \apjl, 594, L83 
\bibitem[Granot \& Taylor(2005)]{gt05} Granot, J., \&
Taylor, G.~B.\ 2005, \apj, 625, 263


\bibitem[Greiner et al.(2001)]{greiner2001a} Greiner, J., Cuby,
J.~G., \& McCaughrean, M.~J.\ 2001, \nat, 414, 522 
\bibitem[Guilbert et al.(1983)]{guil83} Guilbert, P.~W.,
Fabian, A.~C., \& Rees, M.~J.\ 1983, \mnras, 205, 593 



\bibitem[Heger et al.(2005)]{heger05} Heger, A., Woosley,
S.~E., \& Spruit, H.~C.\ 2005, \apj, 626, 350


\bibitem[Hirose et al.(2004)]{hirose04} Hirose, S., Krolik,
J.~H., De Villiers, J., \& Hawley, J.~F.\ 2004, \apj, 606, 1083 
\bibitem[Hjorth et al.(2003)]{hjorth2003} Hjorth, J., et al.\
2003, \nat, 423, 847 
\bibitem[Ho(1999)]{ho99} Ho, L.~C.\ 1999, \apj, 516, 672
\bibitem[Igumenshchev \& Abramowicz (1999)]{ia99} Igumenshchev, I. V., \& Abramowicz, M. A. 1999, \mnras, 303, 309
\bibitem[Igumenshchev \& Abramowicz (2000)]{ia00} Igumenshchev, I. V., \& Abramowicz, M. A. 2000, \apjs, 130, 463
\bibitem[Igumenshchev et al.(2003)]{igumenshchev2003} Igumenshchev,
I.~V., Narayan, R., \& Abramowicz, M.~A.\ 2003, \apj, 592, 1042 
\bibitem[Jorstad et al.(2001)]{jorstad01} Jorstad, S.~G.,
Marscher, A.~P., Mattox, J.~R., Wehrle, A.~E., Bloom, S.~D., \&
Yurchenko, A.~V.\ 2001, \apjs, 134, 181 
\bibitem[Junor et al.(1999)]{junor99} Junor, W., Biretta,
J.~A., \& Livio, M.\ 1999, \nat, 401, 891 
\bibitem[Kaiser et al.(2004)]{kaiser04} Kaiser, C.~R., Gunn,
K.~F., Brocksopp, C., \& Sokoloski, J.~L.\ 2004, \apj, 612, 332 

\bibitem[Kalogera et al.(2001)]{kal01} Kalogera, V., Narayan,
R., Spergel, D.~N., \& Taylor, J.~H.\ 2001, \apj, 556, 340

\bibitem[Kato et al.(2004)]{kato04} Kato, Y., Mineshige, S.,
\& Shibata, K.\ 2004, \apj, 605, 307



\bibitem[Kawabata et al.(2003)]{kawabata2003} Kawabata, K.~S., et
al.\ 2003, \apjl, 593, L19 
\bibitem[Kino \& Takahara(2004)]{kt04} Kino, M., \&
Takahara, F.\ 2004, \mnras, 349, 336 
\bibitem[Khokhlov et al.(1999)]{khok99} Khokhlov, A.~M., H{\"
o}flich, P.~A., Oran, E.~S., Wheeler, J.~C., Wang, L., \&
Chtchelkanova, A.~Y.\ 1999, \apjl, 524, L107 
\bibitem[Kobayashi et al.(1997)]{kob97} Kobayashi, S., Piran,
T., \& Sari, R.\ 1997, \apj, 490, 92


\bibitem[Kohri \& Mineshige(2002)]{kohri2002} Kohri, K., \&
Mineshige, S.\ 2002, \apj, 577, 311 
\bibitem[Kohri et al.(2005)]{kohri2005} Kohri, K., Narayan, R.,
\& Piran, T.\ 2005, astro-ph/0502470 
\bibitem[Koide, Shibata, Kudoh, \& Meier(2002)]{koide2002} Koide,
S., Shibata, K., Kudoh, T., \& Meier, D.~L.\ 2002, Science, 295,
1688 
\bibitem[Komissarov(2004)]{kom04} Komissarov, S.~S.\ 2004,
\mnras, 350, 1431


\bibitem[Komissarov(2005)]{kom05} Komissarov, S.~S.\ 2005,
\mnras, 359, 801
%
%
\bibitem[Konopelko et al.(2003)]{kono03} Konopelko, A.,
Mastichiadis, A., Kirk, J., de Jager, O.~C., \& Stecker, F.~W.\
2003, \apj, 597, 851 
\bibitem[Kouveliotou et al.(1999)]{kouv1999} Kouveliotou, C., et
al.\ 1999, \apjl, 510, L115 
\bibitem[Krawczynski et al.(2002)]{kraw02} Krawczynski, H.,
Coppi, P.~S., \& Aharonian, F.\ 2002, \mnras, 336, 721 
%
\bibitem[Krichbaum et al.(2004)]{krich04} Krichbaum, T.~P., et
al.\ 2004, European VLBI Network on New Developments in VLBI Science
and Technology, 15
%
%
\bibitem[Kulkarni et al.(1998)]{kulkarni1998} Kulkarni, S.~R., et
al.\ 1998, \nat, 395, 663 
\bibitem[Lamb et al.(2004)]{lamb2004} Lamb, D.~Q., Donaghy,
T.~Q., \& Graziani, C.\ 2004, New Astronomy Review, 48, 459 
\bibitem[Lazzati et al.(2004)]{lazzati2004} Lazzati, D., Rossi, E.,
Ghisellini, G., \& Rees, M.~J.\ 2004, \mnras, 347, L1 
\bibitem[Lazzati et al.(2005)]{lazzati2005} Lazzati, D., Begelman,
M., Ghirlanda, G., Ghisellini, G., \& Firmani, C.\ 2005,
astro-ph/0503630 
\bibitem[Leblanc \& Wilson(1970)]{lw70} Leblanc, J.~M.~\&
Wilson, J.~R.\ 1970, \apj, 161, 541 
\bibitem[Leismann et al.(2005)]{leismann2005} Leismann, T.,
Ant{\'o}n, L., Aloy, M.~A., M{\"u}ller, E., Mart{\'{\i}}, J.~M.,
Miralles, J.~A., \& Ib{\'a}{\~n}ez, J.~M.\ 2005, \aap, 436, 503


\bibitem[Levinson \& Blandford(1996)]{lb96} Levinson, A., \&
Blandford, R.\ 1996, \apjl, 456, L29 
\bibitem[Levinson \& Eichler(1993)]{le93} Levinson, A., \&
Eichler, D.\ 1993, \apj, 418, 386 
\bibitem[Levinson \& Eichler(2003)]{le2003} Levinson, A., \& Eichler,
D.\ 2003, \apjl, 594, L19 
\bibitem[Levinson(2005)]{lev05} Levinson, A.\ 2005, astro-ph/0502346
\bibitem[Lewin et al.(1995)]{lewin1995} Lewin, W.~H.~G., van
Paradijs, J., \& van den Heuvel, E.~P.~J.\ 1995, Cambridge
Astrophysics Series, Cambridge, MA: Cambridge University Press,
|c1995, edited by Lewin, Walter H.G.; Van Paradijs, Jan; Van den
Heuvel, Edward P.J., 
\bibitem[Li et al.(1992)]{lcb92} Li, Z., Chiueh, T., \&
Begelman, M.~C.\ 1992, \apj, 394, 459 

\bibitem[Li(1993)]{li1993} Li, Z.-Y.\ 1993, \apj, 415, 118


\bibitem[Li(2000)]{li2000} Li, L.\ 2000, \apjl, 531, L111


\bibitem[Li(2002)]{li2002} Li, L.\ 2002, \apj, 564, 108


\bibitem[Lind et al.(1989)]{lind89} Lind, K.~R., Payne, D.~G.,
Meier, D.~L., \& Blandford, R.~D.\ 1989, \apj, 344, 89




\bibitem[Livio, Ogilvie, \& Pringle(1999)]{lop99} Livio, M.,
Ogilvie, G.~I., \& Pringle, J.~E.\ 1999, \apj, 512, 100 
\bibitem[Lloyd-Ronning et al.(2004)]{lr04} Lloyd-Ronning,
N.~M., Dai, X., \& Zhang, B.\ 2004, \apj, 601, 371 
\bibitem[Lorimer(2001)]{lorimer2001} Lorimer, D.~R.\ 2001, ArXiv
Astrophysics e-prints, arXiv:astro-ph/0104388 
\bibitem[Lovelace(1976)]{lovelace76} Lovelace, R.~V.~E.\ 1976,
\nat, 262, 649 
%
\bibitem[Lyutikov \& Blackman(2001)]{lb01} Lyutikov, M., \&
Blackman, E.~G.\ 2001, \mnras, 321, 177

\bibitem[Lyutikov \& Blandford(2003)]{lb03} Lyutikov, M., \&
Blandford, R.\ 2003, ArXiv Astrophysics e-prints,
arXiv:astro-ph/0312347


\bibitem[Lyutikov, Pariev, \& Blandford(2003)]{lpb03}
Lyutikov, M., Pariev, V.~I., \& Blandford, R.~D.\ 2003, \apj, 597,
998 


\bibitem[MacFadyen \& Woosley(1999)]{mw99} MacFadyen, A. I. \& Woosley, S. E.\
1999, \apj, 524, 262

%
\bibitem[Maraschi \& Tavecchio(2003)]{mt03} Maraschi, L.~\&
Tavecchio, F.\ 2003, \apj, 593, 667 
\bibitem[Martocchia et al.(2002)]{martocchia02} Martocchia, A.,
Matt, G., Karas, V., Belloni, T., \& Feroci, M.\ 2002, \aap, 387,
215 
\bibitem[McClintock \& Remillard(2003)]{mcclintock2003} McClintock,
J.~E., \& Remillard, R.~A.\ 2003, ArXiv Astrophysics e-prints,
arXiv:astro-ph/0306213 
\bibitem[McKinney \& Gammie(2002)]{mg02} McKinney, J.~C., \& Gammie,
C.~F.\ 2002, \apj, 573, 728 
\bibitem[McKinney(2004)]{mckinney2004} McKinney, J.C. 2004, unpublished Ph.D.
        thesis, University of Illinois at Urbana-Champaign, {\rm
        http://rainman.astro.uiuc.edu/$\symbol{126}$jon/research/thesis.pdf}
\bibitem[McKinney \& Gammie(2004)]{mg04} McKinney, J.~C., \&
Gammie, C.~F.\ 2004, \apj, 611, 977 
\bibitem[McKinney(2005a)]{mckinney2005a} McKinney, J.~C.\ 2005a, \apjl, 630, L5

%
\bibitem[McKinney(2005b)]{mckinney2005b} McKinney, J.~C. 2005b, astro-ph/0506368
\bibitem[McKinney(2005c)]{mckinney2005c} McKinney, J.~C. 2005c,
astro-ph/0506369, v1
%
\bibitem[McKinney(2006a)]{mckinney2006a} McKinney, J.~C.\ 2006a, MNRAS, in
press (astro-ph/0601410)


\bibitem[McKinney(2006b)]{mckinney2006b} McKinney, J.~C.\ 2006b, MNRAS, in
press (astro-ph/0601411)

\bibitem[Meier(2003)]{meier2003} Meier, D.~L.\ 2003, New
Astronomy Review, 47, 667 
\bibitem[Meier(2005)]{meier2005} Meier, D.~L.\ 2005, \apss, 300, 55


\bibitem[Merloni et al.(2003)]{merloni2003} Merloni, A., Heinz, S.,
\& di Matteo, T.\ 2003, \mnras, 345, 1057 
\bibitem[Messer et al. and collaborators(1998)]{Trans}
O. E. B. Messer, A. Mezzacappa, S. W. Bruenn, and M. W. Guidry,
Astrophys. J. {\bf 507}, 353 (1998); S. Yamada, H-Th Janka, H.
Suzuki, Astron. Astrophys. {\bf344}, 533 (1999); A. Burrows, T.
Young, P. Pinto, R. Eastman, T. A. Thompson, Astrophys. J. {\bf
539}, 865  (2000); M. Rampp, H-Th Janka, Astrophys. J. {\bf 539},
L33 (2000); M. Liebendoerfer, {\it et al.} Phys. Rev. D {\bf 63},
104003-1 (2001). 
\bibitem[M{\' e}sz{\' a}ros(2002)]{mes02} M{\' e}sz{\' a}ros,
P.\ 2002, \araa, 40, 137 
\bibitem[M{\' e}sz{\' a}ros \& Rees(1997)]{mr97} M{\' e}sz{\' a}ros, P.~\& Rees,M.~J.\ 1997, \apjl, 482, L29
\bibitem[Mirabel et al.(1992)]{mirabel1992} Mirabel, I.~F.,
Rodriguez, L.~F., Cordier, B., Paul, J., \& Lebrun, F.\ 1992, \nat,
358, 215 

\bibitem[Miller-Jones et al.(2006)]{mj06} Miller-Jones,
J.~C.~A., Fender, R.~P., \& Nakar, E.\ 2006, ArXiv Astrophysics
e-prints, arXiv:astro-ph/0601482



\bibitem[Mirabel \& Rodriguez(1994)]{mr94} Mirabel, I.~F.,
\& Rodriguez, L.~F.\ 1994, \nat, 371, 46 
\bibitem[Mirabel \& Rodr{\'{\i}}guez(1999)]{mr99} Mirabel,
I.~F., \& Rodr{\'{\i}}guez, L.~F.\ 1999, \araa, 37, 409 
\bibitem[Misner et al.(1973)]{mtw73} Misner, C.~W., Thorne,
K.~S., \& Wheeler, J.~A.\ 1973, Gravitation, San Francisco:
W.H.~Freeman and Co., 1973


\bibitem[Mizuno et al.(2004)]{mizuno04b} Mizuno, Y., Yamada, S.,
Koide, S., \& Shibata, K.\ 2004, \apj, 615, 389 
\bibitem[Mochkovitch et al.(1993)]{moch93} Mochkovitch, R.,
Hernanz, M., Isern, J., \& Martin, X.\ 1993, \nat, 361, 236

\bibitem[Nakamura et al.(2001)]{nakamura2001} Nakamura, M., Uchida,
Y., \& Hirose, S.\ 2001, New Astronomy, 6, 61 
\bibitem[Nakamura \& Meier(2004)]{nakamura2004} Nakamura, M., \&
Meier, D.~L.\ 2004, \apj, 617, 123 


\bibitem[Narayan et al.(1992)]{narayan1992} Narayan, R., Paczynski,
B., \& Piran, T.\ 1992, \apjl, 395, L83 
\bibitem[Narayan \& Yi(1995)]{ny95} Narayan, R.~\& Yi, I.\
1995, \apj, 452, 710 
\bibitem[Narayan et al.(2001)]{narayan2001} Narayan, R., Piran, T.,
\& Kumar, P.\ 2001, \apj, 557, 949 
\bibitem[Narayan et al.(2003)]{narayan2003} Narayan, R.,
Igumenshchev, I.~V., \& Abramowicz, M.~A.\ 2003, \pasj, 55, L69 
\bibitem[Noble et al.(2005)]{noble05} Noble, S.C. , Gammie, C.F., McKinney,
  J.C., Del Zanna, L.D., 2005, \apj, accepted, astro-ph/0512420
\bibitem[Okamoto(1999)]{okamoto1999} Okamoto, I.\ 1999, \mnras,
307, 253 
\bibitem[Okamoto(2000)]{okamoto2000} Okamoto, I.\ 2000, \mnras,
318, 250

%
\bibitem[Ouyed et al.(2003)]{ocp03} Ouyed, R., Clarke, D.~A.,
\& Pudritz, R.~E.\ 2003, \apj, 582, 292


\bibitem[Paczynski(1998)]{pac98} Paczynski, B.\ 1998, \apjl,
494, L45 
\bibitem[Papaloizou \& Pringle(1983)]{pp83} Papaloizou,
J.~C.~B.~\& Pringle, J.~E.\ 1983, \mnras, 202, 1181 
\bibitem[Phinney(1983)]{phi83} Phinney, E.S. 1983, unpublished Ph.D.
        thesis, Cambridge University



\bibitem[Piran(2005)]{piran2005} Piran, T.\ 2005, Reviews of
Modern Physics, 76, 1143 
\bibitem[Popham et al.(1999)]{pwf99} Popham, R., Woosley,
S.~E., \& Fryer, C.\ 1999, \apj, 518, 356 
\bibitem[Proga et al.(2003)]{proga03} Proga, D., MacFadyen,
A.~I., Armitage, P.~J., \& Begelman, M.~C.\ 2003, \apjl, 599, L5


%
\bibitem[Punsly \& Coroniti(1990a)]{pc90a} Punsly, B., \&
Coroniti, F.~V.\ 1990, \apj, 354, 583
%
\bibitem[Punsly \& Coroniti(1990b)]{pc90b} Punsly, B., \&
Coroniti, F.~V.\ 1990, \apj, 350, 518
%
%
\bibitem[Punsly(2001)]{punsly2001} Punsly, B.\ 2001, Black hole
gravitohydromagnetics, Berlin: Springer, 2001, xii, 400 p.,
Astronomy and astrophysics library, ISBN 3540414665, 
\bibitem[Ramirez-Ruiz \& Socrates(2005)]{rs2005} Ramirez-Ruiz,
E., \& Socrates, A.\ 2005, ArXiv Astrophysics e-prints,
arXiv:astro-ph/0504257 
\bibitem[Rees et al.(1982)]{rees82} Rees, M.~J., Phinney,
E.~S., Begelman, M.~C., \& Blandford, R.~D.\ 1982, \nat, 295, 17 
\bibitem[Reynolds et al.(1996)]{reynolds96} Reynolds, C.~S., di
Matteo, T., Fabian, A.~C., Hwang, U., \& Canizares, C.~R.\ 1996,
\mnras, 283, L111 

\bibitem[Rosen et al.(1999)]{rosen99} Rosen, A., Hardee, P.~E.,
Clarke, D.~A., \& Johnson, A.\ 1999, \apj, 510, 136


\bibitem[Salpeter(1964)]{salpeter64} Salpeter, E. E. 1964, \apj, 140, 796

\bibitem[Sauty et al.(2002)]{sauty2002} Sauty, C., Tsinganos, K.,
\& Trussoni, E.\ 2002, LNP Vol.~589: Relativistic Flows in
Astrophysics, 589, 41


\bibitem[Scheck et al.(2002)]{scheck2002} Scheck, L., Aloy, M.~A.,
Mart{\'{\i}}, J.~M., G{\' o}mez, J.~L., M\"{u}ller, E.\ 2002,
\mnras, 331, 615 

\bibitem[Shay et al.(2001)]{shay01} Shay, M.~A., Drake, J.~F.,
Rogers, B.~N., \& Denton, R.~E.\ 2001, \jgr, 106, 3759




\bibitem[Shu(1997)]{shu97} Shu, C.W., 1997, NASA/CR-97-206253, No.97-65

%
%
\bibitem[Sikora et al.(2003)]{sbcp03} Sikora, M., Begelman, M. C., Coppi, P., \& Proga, D. 2003, \apj, submitted, astro-ph/0309504
\bibitem[Sikora et al.(2005)]{sikora2005} Sikora, M., Begelman, M.~C.,
Madejski, G.~M., \& Lasota, J.\ 2005, \apj, 625, 72 
\bibitem[Sikora \& Madejski(2000)]{sm00} Sikora, M., \&
Madejski, G.\ 2000, \apj, 534, 109 
\bibitem[Soderberg \& Ramirez-Ruiz(2003)]{sr03} Soderberg,
A.~M.~\& Ramirez-Ruiz, E.\ 2003, AIP Conf.~Proc.~662: Gamma-Ray
Burst and Afterglow Astronomy 2001: A Workshop Celebrating the First
Year of the HETE Mission, 662, 172 
\bibitem[Springel et al.(2004)]{springel2004} Springel, V., Di
Matteo, T., \& Hernquist, L.\ 2004, ArXiv Astrophysics e-prints,
arXiv:astro-ph/0411108 
\bibitem[Spruit \& Taam(1990)]{st90} Spruit, H.~C., \& Taam,
R.~E.\ 1990, \aap, 229, 475


\bibitem[Spruit, Daigne, \& Drenkhahn(2001)]{sdd01} Spruit, H. C., Daigne, F., \& Drenkhahn, G. 2001, \aap, 369, 694




\bibitem[Stergioulas(2003)]{st03} Stergioulas, N.\ 2003,
Living Reviews in Relativity, 6, 3 
\bibitem[Stone et al. (1999)]{spb99} Stone, J. M., Pringle, J. E., \& Begelman,
 M. C. 1999, \mnras, 310, 1002
\bibitem[Subramanian et al.(1999)]{sub99} Subramanian, P.,
Becker, P.~A., \& Kazanas, D.\ 1999, \apj, 523, 203

\bibitem[Symbalisty(1984)]{sym84} Symbalisty, E.~M.~D.\ 1984,
\apj, 285, 729 
\bibitem[Stanek et al.(2003)]{stanek2003} Stanek, K.~Z., et al.\
2003, \apjl, 591, L17 
\bibitem[Taylor et al.(2004a)]{taylor2004a} Taylor, G.~B., Frail,
D.~A., Berger, E., \& Kulkarni, S.~R.\ 2004, \apjl, 609, L1 
\bibitem[Taylor et al.(2004b)]{taylor2004b} Taylor, G.~B., Momjian,
E., Pihlstrom, Y., Ghosh, T., \& Salter, C.\ 2004, ArXiv
Astrophysics e-prints, arXiv:astro-ph/0412483 
\bibitem[Takahashi et al.(1990)]{tak90} Takahashi, M. , Nitta, S. ,
        Tatematsu, Y.  \& Tomimatsu, A.  1990, \apj, 363, 206
\bibitem[Tanaka et al.(1995)]{tanaka1995} Tanaka, Y., et al.\
1995, \nat, 375, 659 
\bibitem[Thompson(1994)]{thompson94} Thompson, C.\ 1994, \mnras, 270, 480
\bibitem[Thompson \& Duncan(1995)]{td95} Thompson, C., \&
Duncan, R.~C.\ 1995, \mnras, 275, 255 
\bibitem[Thompson \& Duncan(1996)]{td96} Thompson, C., \&
Duncan, R.~C.\ 1996, \apj, 473, 322 
\bibitem[Thorsett \& Chakrabarty(1999)]{tc99} Thorsett,
S.~E., \& Chakrabarty, D.\ 1999, \apj, 512, 288 
%
\bibitem[Tomimatsu(1994)]{tomimatsu1994} Tomimatsu, A.\ 1994, \pasj,
46, 123


\bibitem[Tomimatsu et al.(2001)]{tom01} Tomimatsu, A.,
Matsuoka, T., \& Takahashi, M.\ 2001, \prd, 64, 123003


\bibitem[Tomimatsu \& Takahashi(2003)]{tt03} Tomimatsu, A.,
\& Takahashi, M.\ 2003, \apj, 592, 321


\bibitem[Tsinganos \& Bogovalov(2005)]{tb05} Tsinganos, K.,
\& Bogovalov, S.\ 2005, AIP Conf.~Proc.~745: High Energy Gamma-Ray
Astronomy, 745, 148 
\bibitem[Uemura et al.(2003)]{uemura2003} Uemura, M., et al.\
2003, \nat, 423, 843 
\bibitem[Urry \& Padovani(1995)]{up95} Urry, C.~M., \&
Padovani, P.\ 1995, \pasp, 107, 803 
\bibitem[Uzdensky(2005)]{uzdensky2005} Uzdensky, D.~A.\ 2005, \apj,
620, 889 
\bibitem[Spruit \& Uzdensky(2005)]{pu05} Spruit, H.~C., \&
Uzdensky, D.~A.\ 2005, \apj, 629, 960


\bibitem[Vlahakis(2004)]{vlah04} Vlahakis, N.\ 2004, \apss, 293, 67
\bibitem[Wang \& Wheeler(1996)]{ww96} Wang, L.~\& Wheeler,
J.~C.\ 1996, \apjl, 462, L27 
\bibitem[Wang, Howell, H{\" o}flich, \& Wheeler(2001)]{wang01}
Wang, L., Howell, D.~A., H{\" o}flich, P., \& Wheeler, J.~C.\ 2001,
\apj, 550, 1030 
\bibitem[Wang et al.(2002)]{wang02} Wang, L., et al.\ 2002,
\apj, 579, 671 
\bibitem[Wang, Baade, H{\" o}flich, \& Wheeler(2003)]{wang03}
Wang, L., Baade, D., H{\" o}flich, P., \& Wheeler, J.~C.\ 2003,
\apj, 592, 457 
\bibitem[Wheeler, Yi, H{\" o}flich, \& Wang(2000)]{wyhw00}
Wheeler, J.~C., Yi, I., H{\" o}flich, P., \& Wang, L.\ 2000, \apj,
537, 810 
\bibitem[Woosley(1993)]{w93} Woosley, S.~E.\ 1993, \apj,
405, 273 
\bibitem[Woosley \& Weaver(1995)]{ww95} Woosley, S.~E., \&
Weaver, T.~A.\ 1995, \apjs, 101, 181 
\bibitem[Woosley \& Weaver(1986)]{ww86} Woosley, S.~E.~\&
Weaver, T.~A.\ 1986, \araa, 24, 205 
\bibitem[Woosley et al.(2003)]{woosley03} Woosley, S.~E., Zhang,
W., \& Heger, A.\ 2003, From Twilight to Highlight: The Physics of
Supernovae, 87


\bibitem[Zel'dovich(1964)]{z64} Zel'dovich Y. B. 1964,
  Sov. Phys.-Dokl., 9, 195
\bibitem[Zhang B. et al.(2004)]{zdlm04} Zhang, B., Dai, X.,
Lloyd-Ronning, N.~M., \& M{\' e}sz{\' a}ros, P.\ 2004, \apjl, 601,
L119 
\bibitem[Zhang, Woosley, \& MacFadyen(2003)]{zwm03} Zhang, W., Woosley, S. E., \& MacFadyen, A. I.  2003, \apj, 586, 356
\bibitem[Zhang W. et al.(2004)]{zwh04} Zhang, W., Woosley,
S.~E., \& Heger, A.\ 2004, \apj, 608, 365 
\bibitem[Zhang \& MacFadyen(2005)]{zm05} Zhang, W., \&
MacFadyen, A.~I.\ 2005, astro-ph/0505481 
%



\end{thebibliography}
\end{document}